\documentclass[english,aps,preprint,prb]{revtex4}
\usepackage[T1]{fontenc}
\usepackage[latin1]{inputenc}
\usepackage{amsmath}
\usepackage{graphicx}
\usepackage{amssymb}

\makeatletter
\usepackage{babel}
\makeatother
\begin{document}

\preprint{This line only printed with preprint option}

\title{Exact Green's functions and Bosonization of a Luttinger liquid coupled
to impedances.}

\author{Khuôn-Viêt Pham}

\email{pham@lps.u-psud.fr}

\affiliation{Laboratoire de Physique des Solides, Université Paris-Sud, Orsay,
France.}

\begin{abstract}
The exact Green's functions of a Luttinger Liquid (LL) connected to
impedances are computed at zero and finite temperature. It is also
shown that if the resistances are equal to the characteristic impedance
of the Luttinger liquid then the finite Luttinger liquid connected
to resistors is equivalent to an infinite Luttinger liquid. Impedance
boundary conditions (IBC) include also as a special limit the case
of open boundary conditions, which are explicitly recovered. Finally
bosonization for a LL with IBC is proven to hold.
\end{abstract}
\maketitle

\section{introduction}

Bosonization is one of the standard methods for one dimensional quantum
field theories \cite{mycitation}. Discovered independently in condensed
matter and high-energy physics it is the main tool which has allowed
to construct the concept of 'Luttinger liquid' inspired by the physics
of the Tomonaga and Luttinger Hamiltonians \cite{key-2}. The LL is
a universality class of 1D critical systems comprising models as important
as the Heisenberg spin chain, the Hubbard model or the Calogero-Sutherland
model. 

Various boundary conditions (abbreviated as BC throughout the paper)
have been considered in the past for a LL. The earliest studies focused
on the infinite system and the finite-size LL with periodic boundary
conditions (PBC). The properties in both cases can be related through
a conformal transformation \cite{mycitation}. In particular the finite-size
properties have been very useful in conjunction with numerics for
extraction of the LL parameters. Later more general boundary conditions
were also considered: for instance twisted boundary conditions (TBC)
or open boundary conditions (OBC). TBC led to the discovery of even-odd
effects in a LL and periodicity of permanent currents with the flux
\cite{key-3}. The OBC which allow description of broken chains were
found to have a dramatic effect on critical exponents leading to boundary
exponents in addition to bulk ones and bridging the physics of the
LL to that of boundary conformal field theory \cite{key-4,key-5}.
More recently dissipative boundary conditions were also introduced
for the LL and were used to compute transport properties of the LL:
they describe the coupling of a LL wire to electrodes; they comprise
the so-called 'radiative boundary conditions' (which relate time and
space derivative of the boson fields \cite{key-6}) and a 'chemical
potential matching boundary conditions' \cite{key-7} . Other dissipative
boundary conditions include the 'Impedance Boundary Conditions' (IBC)
introduced by the author \cite{ac}: they consist in a LL connected
at its boundaries to two impedances. The IBC can actually be shown
to encompass both 'radiative boundary conditions' and 'chemical potential
matching boundary conditions' which constitute special cases of the
IBC with boundary impedances set at half-a-quantum of resistance $h/2e^{2}$.
\cite{key-9}

We will deal in the present paper with the following issues for a
LL with IBC: (1) computing its exact Green's functions and two-point
correlators; this will pave the groundwork allowing for (2) extending
the usual bosonization technique to such a dissipative system.

Indeed there is no reason to believe that bosonization of such a system
is a valid procedure: bosonization is well established in the non-dissipative
situations (infinite LL, finite-size LL with PBC, TBC or OBC) but
a LL connected to resistors is a dissipative system: plasmons have
now a finite-lifetime. Actually to the author's knowledge bosonization
has not been shown to be valid for any LL with dissipative boundary
condition. One strategy to show the validity of bosonization is to
start from fermions and then (by considering the anomalous current
algebra of density operators) to transform the fermionic Hamiltonian
into a bosonic one \cite{mycitation}. Such a course is in our case
plagued with difficulties related to dissipation: the density eigenmodes
of the cavity do not form a neat orthogonal basis of states and do
not quantize as free bosons. Such problems are actually symptoms of
non-hermitian Hamiltonian physics and the existence of non-trivial
self-energies: this is a recurrent issue for open systems which is
well-known and has led to recourse to biorthogonal bases of states
in such diverse contexts as mesoscopic transport \cite{key-11}, laser
physics (leaking cavities in QED where quantization of the gauge field
in terms of photons breaks down) \cite{key-12}, acoustics \cite{key-13},
black hole physics \cite{key-14}, etc. Biorthogonal bases of states
lead however for the bosonization program to unnecessary complications.

Nevertheless it will be shown that bosonization does hold for the
model at hand (LL with IBC). Instead of following the afore mentioned
strategy for bosonization we will find more convenient to start from
a bosonic theory and then fermionize it. As an interesting side result
of our proof we will compute the exact Green's functions and correlators
of the boson Hamiltonian. Another interesting side result with potential
applications is the finding that with suitably chosen resistances
the finite LL with impedance boundary conditions (IBC) is equivalent
to an infinite LL (namely by identity of their Green's functions):
in a mesoscopic setting the LL in principle can not be abstracted
from its surroundings so that the intrinsic properties of an infinite
LL are not directly accessible. We show a conceptually simple way
out which exploits the fact that a LL can be viewed as a quantum transmission
line.

OBC will also be shown to be a special case of IBC corresponding to
infinite resistances.

The paper structure will be as follows: 

- (1) in section \ref{sec:model} we introduce a model of a (fermionic)
LL connected to resistors through boundary conditions (impedance boundary
conditions IBC); these boundary conditions are then recast equivalently
in terms of boundary conditions for chiral bosons.

- (2) In the next two sections \ref{sec:Green's-function-of} and
\ref{sec:Two-point-Correlators-of} we next focus on the bosonic model
and compute exactly its Green's functions and correlators (at zero
and finite temperature). As side results we obtain the exact finite-frequency
conductivity of the LL with IBC and we prove the equivalence of an
infinite LL to a finite-size LL with IBC with suitably chosen resistances.

- (3) We prove bosonization for our model in section \ref{sec:Bosonization.}.

- (4) We compute the fermion correlation functions in \ref{sec:Fermion-correlators.}.

Finally an Appendix (Appendix B) is devoted to the single issue of
recovering explicitly the fermionic Open BC and the Green's function,
with results identical to those published in the litterature.

\section{Model\label{sec:model}}

\subsection{Notations and definitions.}

\textbf{Phase fields:} We consider throughout the paper the standard
LL Hamiltonian which written in terms of the usual phase fields reads:

\[
H=\frac{\hbar u}{2}\int_{-L/2}^{L/2}dx\,\,\frac{1}{K}\:\left(\partial_{x}\phi\right)^{2}+K\:\Pi^{2}\]

where the fields $\phi$ and $\Pi$ are canonical conjugates \[
\left[\phi(x,t);\,\Pi(y,t)\right]=i\,\delta(x-y).\]
In terms of fermionic operators the fermion density operator is related
to $\phi$ through:

\[
\rho-\rho_{0}=\frac{1}{\sqrt{\pi}}\:\partial_{x}\phi.\]

(Of course such an identification is not fully warranted at this stage
but we will show later it does hold even for a LL connected to resistors;
for the time being we may view the relations as \emph{defining} abstractly
the operator $\rho$ rather than equating it with the operator $\psi^{+}\psi$.
Similar remarks apply for all the operators defined for fermions such
as current, etc.)

The phase field $\Theta$ is defined per:\[
\Pi=-\partial_{x}\Theta\]
and \[
\left[\phi(x,t);\,\Theta(y,t)\right]=i\,\theta(x-y)\]
where $\theta$ is the Heaviside step function.

\textbf{Chiral fields:} The equations of motion of the phase fields:
\begin{eqnarray*}
\partial_{t}\phi & = & -uK\:\partial_{x}\Theta\\
\partial_{t}\Theta & = & -\frac{u}{K}\:\partial_{x}\phi\end{eqnarray*}
imply:\[
\partial_{x}\left(\phi\pm K\Theta\right)=\mp\frac{1}{u}\,\partial_{t}\left(\phi\pm K\Theta\right).\]

The fields $\phi\pm K\Theta$ are therefore chiral and we define chiral
phase fields and chiral densities:

\begin{eqnarray*}
\phi_{\pm} & = & \frac{\phi\pm K\Theta}{2},\\
\rho_{\pm} & = & \frac{1}{\sqrt{\pi}}\,\partial_{x}\phi_{\pm}.\end{eqnarray*}

They obey: $\phi_{\pm}(x,t)=\phi_{\pm}(x\mp ut)$. Evidently $\rho-\rho_{0}=\rho_{+}+\rho_{-}$.

\textbf{Current:} The particle current density is:

\begin{eqnarray*}
i & = & -\frac{1}{\sqrt{\pi}}\:\partial_{t}\phi\\
 & = & \frac{uK}{\sqrt{\pi}}\:\partial_{x}\Theta\\
 & = & u\:(\rho_{+}-\rho_{-})\end{eqnarray*}
where the first line follows from current conservation and the others
from the equations of motion of the phase fields.

\textbf{Chiral chemical potentials:}

We define the following operators $\mu_{\pm}$ (they will prove convenient
to define our model):\[
\mu_{\pm}(x,t)=\frac{\delta H}{\delta\rho_{\pm}(x,t)}\]
 where functional differentiation with respect to the particle density
$\rho_{\pm}(x,t)$ has been performed. Physically they correspond
to chemical potential operators: their average value yields the energy
needed to add one particle at position $x$ to the chiral density:
$\rho_{\pm}\longrightarrow\rho_{\pm}+\delta(x)$. These chiral chemical
potentials correspond to the plasma chiral eigenmodes of the Luttinger
liquid and \textit{not to the left or right moving (bare) electrons}.
An average chemical potential can be defined also as:

\[
\mu=\frac{\mu_{+}+\mu_{-}}{2}.\]

From their definition it follows that: \begin{equation}
\mu_{\pm}=\frac{hu}{K}\,\rho_{\pm},\end{equation}
where we used the relation: \[
H=\frac{\pi\hbar u}{K}\int_{-L/2}^{L/2}dx\:\rho_{+}^{2}+\rho_{-}^{2}.\]
 Therefore the electrical current $i_{e}=e\: i$ :

\begin{eqnarray}
i_{e}(x,t) & = & K\frac{e}{h}\left(\mu_{+}-\mu_{-}\right)\nonumber \\
 & = & \frac{1}{2Z_{0}}\,\left(\frac{\mu_{+}}{e}-\frac{\mu_{-}}{e}\right)\label{currentb}\end{eqnarray}
where we have introduced the \textbf{characteristic impedance} of
the LL:

\[
Z_{0}=\frac{h}{2Ke^{2}}.\]

As explained in \cite{ac} the LL Hamiltonian is identical with that
of a quantum transmission line with a characteristic impedance $Z_{0}$
given as above.

\subsection{Model.}

Our model consists in a LL connected in series with two impedances.
We will use a description of the LL connected to reservoirs \cite{ac}
which is the exact implementation of the load impedances boundary
conditions customary for transmission lines or sound waves in tubes. 

We thus assume the following boundary conditions:

\begin{eqnarray}
Z_{S}\: i_{e}(-L/2,t) & = & V_{S}(t)-\frac{\mu(-L/2,t)}{e},\label{bound1}\\
Z_{D}\: i_{e}(L/2,t) & = & \frac{\mu(L/2,t)}{e}-V_{D}(t).\nonumber \end{eqnarray}

\begin{figure}
\includegraphics{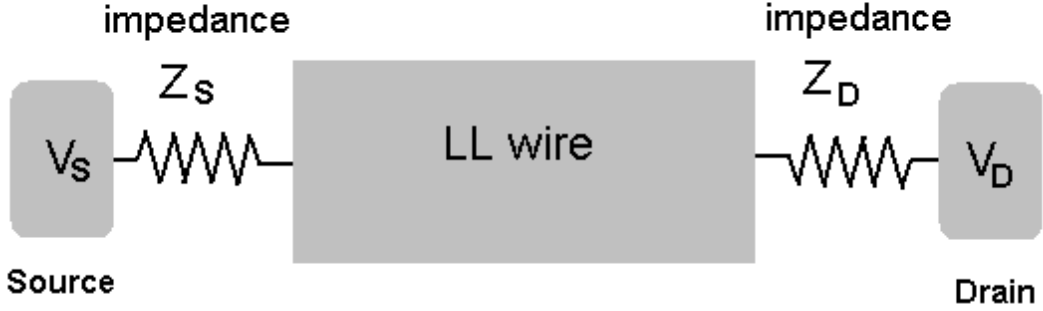}

\caption{\label{fig-formalisme}Impedance boundary conditions: the LL wire
is connected to two electrodes at voltages $V_{S}$ and $V_{D}$ through
two impedances.}
\end{figure}

$Z_{S}$ and $Z_{D}$ are interface impedances (at respectively the
source and the drain) \textit{which for simplicity will be assumed
to be positive real numbers} throughout the paper (in other words
they represent resistors; but more general situations could be discussed
with complex impedances, which is why we stick in this paper to viewing
them as impedances). $i_{e}(x,t)$ is the current operator, and source
and drain are set at a voltage $V_{S}$ or $V_{D}$ (see Fig.\ref{fig-formalisme}).
The Heisenberg picture is assumed so that we work with time-dependent
operators.

As one can see the boundary conditions are tantamount to assuming
Ohm's law at the boundaries of the system $U=RI$: the current is
proportional to a voltage drop between the reservoir and the LL wire
and the proportionality constant is just a resistance. In the following
the source and drain voltages will be set to zero since we want to
compute the equilibrium Green's function (in the absence of external
voltage). 

It is instructive to recast the (equilibrium) boundary conditions
in terms of the chiral densities. This yields:

\begin{eqnarray*}
Z_{S}\: u\, e\:\left[\rho_{+}(-L/2)-\rho_{-}(-L/2)\right] & = & -\frac{hu}{2Ke}\:\left[\rho_{+}(-L/2)+\rho_{-}(-L/2)\right],\\
Z_{D}\: u\, e\:\left[\rho_{+}(L/2)-\rho_{-}(L/2)\right] & = & \frac{hu}{2Ke}\:\left[\rho_{+}(L/2)+\rho_{-}(L/2)\right].\end{eqnarray*}
So that:\begin{eqnarray}
\rho_{+}(-L/2) & = & \frac{Z_{S}-Z_{0}}{Z_{S}+Z_{0}}\:\rho_{-}(-L/2),\label{coef}\\
\rho_{-}(L/2) & = & \frac{Z_{D}-Z_{0}}{Z_{D}+Z_{0}}\:\rho_{+}(L/2).\nonumber \end{eqnarray}
This introduces reflection coefficients for the density: \begin{equation}
r_{S}=\frac{Z_{S}-Z_{0}}{Z_{S}+Z_{0}},\quad r_{D}=\frac{Z_{D}-Z_{0}}{Z_{D}+Z_{0}}\text{.}\label{reflec}\end{equation}
These expressions deserve some comment: they are just what one would
expect for a classical transmission line connected to load and drain
impedances. The (classical) equations of motion are indeed also valid
at the quantum level since the LL Hamiltonian is quadratic so we might
have anticipated any linear relation to carry on. The basic physics
of the boundary conditions considered in this paper are therefore
those of standing waves in a transmission line produced by reflections
at the boundaries due to impedance mismatch. 

The reflection coefficients for the phase fields can also be derived;
from eq.(\ref{coef}) it follows:\begin{eqnarray}
\partial_{x}\phi_{+}(-L/2,t) & = & r_{S}\:\partial_{x}\phi_{-}(-L/2,t),\label{eq:bound2}\\
\partial_{x}\phi_{-}(L/2,t) & = & r_{D}\:\partial_{x}\phi_{+}(L/2,t).\nonumber \end{eqnarray}

The IBC can then be rewritten as conditions on the non-chiral phase
field:

\begin{eqnarray}
-Z_{S}\:\frac{1}{u}\partial_{t}\phi(-L/2,t) & = & Z_{0}\partial_{x}\phi(-L/2,t),\label{boundprime}\\
Z_{D}\:\frac{1}{u}\partial_{t}\phi(-L/2,t) & = & Z_{0}\partial_{x}\phi(-L/2,t).\nonumber \end{eqnarray}

As an aside we note that open boundary conditions (OBC) are recovered
by setting $r_{S}=r_{D}=1$ (see Appendix B).

\section{Green's function of the phase field.\label{sec:Green's-function-of}}

\subsection{Green's function of the phase field $\phi$.}

Having derived boundary conditions for the boson fields we now forget
the underlying fermions and focus on the following model:

\[
H=\frac{\hbar u}{2}\int_{-L/2}^{L/2}dx\,\,\frac{1}{K}\:\left(\partial_{x}\phi\right)^{2}+K\:\Pi^{2}\]

where as before the fields $\phi$ and $\Pi$ are canonical conjugates
and: \[
\phi_{\pm}=\frac{\phi\pm K\Theta}{2}.\]

The problem we will tackle is the following: solving for the retarded
Green's function of the bosonic Hamiltonian subjected to the boundary
conditions for the fields:\begin{eqnarray*}
\partial_{x}\phi_{+}(-L/2,t) & = & r_{S}\:\partial_{x}\phi_{-}(-L/2,t),\\
\partial_{x}\phi_{-}(L/2,t) & = & r_{D}\:\partial_{x}\phi_{+}(L/2,t).\end{eqnarray*}
These boundary conditions introduce dissipation in the problem.

The Green's function can be conveniently divided into four chiral
components: $G_{R}=G_{R}^{++}+G_{R}^{--}+G_{R}^{+-}+G_{R}^{-+}$ where
$G_{R}^{\pm\pm}(x,t;y,t')=-i\,\theta(t-t')\left\langle \left[\phi_{\pm}(x,t),\phi_{\pm}(y,t')\right]\right\rangle $
for the retarded Green's function and appropriate definitions for
the advanced Green's function. Since we will also need the chiral
Green's functions for the bosonization proof instead of directly computing
the full Green's function we will first compute the chiral Green's
functions.

Using the equal-time commutation relations for the chiral fields, 

\begin{eqnarray*}
\left[\phi_{\pm}(x,t),\phi_{\pm}(y,t)\right] & = & \pm\frac{iK}{4}\: sgn(x-y),\\
\left[\phi_{+}(x,t),\phi_{-}(y,t)\right] & =- & \frac{iK}{4},\end{eqnarray*}
one gets the equations of motion for the chiral Green's functions
as: \[
\left[\frac{\partial}{\partial x}\pm\frac{1}{u}\frac{\partial}{\partial t}\right]G^{\pm\pm}(x,t;y,0)=\frac{K}{4u}\, sgn(x-y)\,\delta(t),\]
and: \[
\left[\frac{\partial}{\partial x}\pm\frac{1}{u}\frac{\partial}{\partial t}\right]G^{\pm\mp}(x,t;y,0)=-\frac{K}{4u}\,\delta(t).\]
The impedance boundary conditions imply:

\begin{eqnarray*}
\partial_{x}G_{R}^{++} & =r_{S}\partial_{x}G_{R}^{-+} & ,\; x=-L/2,\\
\partial_{x}G_{R}^{+-} & =r_{S}\partial_{x}G_{R}^{--} & ,\; x=-L/2,\\
\partial_{x}G_{R}^{--} & =r_{D}\partial_{x}G_{R}^{+-} & ,\; x=L/2,\\
\partial_{x}G_{R}^{-+} & =r_{D}\partial_{x}G_{R}^{++} & ,\; x=L/2.\end{eqnarray*}
After Fourier transforming according to:\[
f(t)=\int_{-\infty}^{\infty}\frac{d\omega}{2\pi}\, f(\omega)\, e^{-i\omega t},\]
and defining $k=\omega/u$ the equations of motion imply the following
forms: \begin{eqnarray*}
G^{++} & = & \theta(x-y)\left(\frac{a^{>}}{ik}\, e^{ikx}-\frac{K}{4i\omega}\right)+\theta(y-x)\left(\frac{a^{<}}{ik}\, e^{ikx}+\frac{K}{4i\omega}\right),\\
G^{--} & = & \theta(x-y)\left(-\frac{b^{>}}{ik}\, e^{-ikx}+\frac{K}{4i\omega}\right)+\theta(y-x)\left(-\frac{b^{<}}{ik}\, e^{-ikx}-\frac{K}{4i\omega}\right).\end{eqnarray*}
Likewise:\begin{eqnarray*}
G^{+-} & = & c\, e^{ikx}+\frac{K}{4i\omega},\\
G^{-+} & = & d\, e^{-ikx}-\frac{K}{4i\omega}.\end{eqnarray*}
The boundary conditions imply the following relations:\begin{eqnarray*}
a^{<} & = & -ik\, r_{S}\, e^{i\varphi}\, d\\
a^{>} & = & -ik\, r_{D}^{-1}\, e^{-i\varphi}\, d\\
b^{<} & = & ik\, r_{S}^{-1}\, e^{-i\varphi}\, c\\
b^{>} & = & ik\, r_{D}\, e^{i\varphi}\, c.\end{eqnarray*}
where we have defined a phase $\varphi$ corresponding to the phase
accumulated along the wire by the plasma wave:\[
\varphi=kL.\]
Using now the discontinuity of the derivatives of $G^{++}$ and $G^{--}$yields:\begin{eqnarray*}
a^{>}-a^{<} & = & \frac{K}{2u}\, e^{-iky},\\
b^{>}-b^{<} & = & \frac{K}{2u}\, e^{iky}.\end{eqnarray*}
Finally: \begin{eqnarray*}
G_{R}^{++}(x,y,\omega) & = & \frac{K}{2i\omega\left(1-r_{S}r_{D}e^{i2\varphi}\right)}\, e^{i\frac{\omega}{u}(x-y)}\,\left[\theta(x-y)+\theta(y-x)\, r_{S}r_{D}\, e^{i2\varphi}\right]\\
 &  & -\frac{K}{4i\omega}\, sgn(x-y),\\
G_{R}^{--}(x,y,\omega) & = & \frac{K}{2i\omega\left(1-r_{S}r_{D}e^{i2\varphi}\right)}\, e^{-i\frac{\omega}{u}(x-y)}\,\left[\theta(y-x)+\theta(x-y)\, r_{S}r_{D}\, e^{i2\varphi}\right]\\
 &  & +\frac{K}{4i\omega}\, sgn(x-y),\\
G_{R}^{+-}(x,y,\omega) & =- & \frac{K}{2i\omega}\,\frac{r_{S}\, e^{i\varphi}}{\left(1-r_{S}r_{D}e^{i2\varphi}\right)}\, e^{i\frac{\omega}{u}(x+y)}+\frac{K}{4i\omega},\\
G_{R}^{-+}(x,y,\omega) & =- & \frac{K}{2i\omega}\,\frac{r_{D}\, e^{i\varphi}}{\left(1-r_{S}r_{D}e^{i2\varphi}\right)}\, e^{-i\frac{\omega}{u}(x+y)}-\frac{K}{4i\omega}.\end{eqnarray*}

Gathering all terms the full Green's function is thus:\begin{eqnarray}
G_{R}(x,y,\omega)=\frac{K}{2i(\omega+i\delta)}\frac{1}{1-r_{S}r_{D}e^{i2\varphi}}\label{eq:green}\\
\times\left\{ \theta(x-y)\left[e^{i\frac{\omega}{u}(x-y)}+r_{S}r_{D}e^{-i\frac{\omega}{u}(x-y-2L)}\right]\right. & +\theta(y-x)\left[e^{-i\frac{\omega}{u}(x-y)}+r_{S}r_{D}e^{i\frac{\omega}{u}(x-y+2L)}\right]\nonumber \\
-\left.r_{S}\, e^{i\frac{\omega}{u}(x+y+L)}-r_{D}\, e^{i\frac{\omega}{u}(-x-y+L)}\right\} \nonumber \end{eqnarray}

(where $\varphi=kL$). As a function of time this yields:\begin{eqnarray*}
G_{R}(x,t;y,0)= & -\frac{K}{2}\sum_{n=0}^{\infty}(r_{S}r_{D})^{n}\left\{ \theta\left(t-\frac{2nL+\left|x-y\right|}{u}\right)+r_{S}r_{D}\theta\left(t-\frac{2(n+1)L-\left|x-y\right|}{u}\right)\right.\\
 & -\left.r_{S}\,\theta\left(t-\frac{(2n+1)L+x+y}{u}\right)-r_{D}\,\theta\left(t-\frac{(2n+1)L-x-y}{u}\right)\right\} \end{eqnarray*}
For $t<0$, $G_{R}=0$ as it should be.

\textbf{Technical note:} the studious reader interested in deriving
directly the advanced Green's function from the boundary conditions
should take note that they must be modified. The reason is simple:
a retarded Green's function which describes outgoing waves (away from
the system) are clearly compatible with dissipative BC; but an advanced
Green's function describes incoming waves. So we need BC invariant
under time-reversal: to render the time derivative compatible with
time-reversal we multiply it by $sgn(t)$ \begin{eqnarray}
-sgn(t)\: Z_{S}\:\partial_{t}\phi(-L/2,\: t) & = & Z_{0}\partial_{x}\phi(-L/2,t),\label{b}\\
sgn(t)\: Z_{D}\:\partial_{t}\phi(L/2,\: t) & = & Z_{0}\partial_{x}\phi(L/2,\: t).\nonumber \end{eqnarray}
In terms of the chiral fields this implies: \begin{eqnarray*}
\partial_{x}\phi_{+}(-L/2,t) & = & r_{S}^{sgn(t)}\:\partial_{x}\phi_{-}(-L/2,t),\\
\partial_{x}\phi_{-}(L/2,t) & = & r_{D}^{sgn(t)}\:\partial_{x}\phi_{+}(L/2,t).\end{eqnarray*}

One can check that the BC are now compatible with the usual relation
\[
G_{A}(x,t;\, y,0)=G_{R}(y,-t;\, x,0).\]

Note also that the two-point correlators mix advanced and retarded
Green's functions: therefore they will obey these modified boundary
conditions as can be readily checked.

\subsection{Discussion.}

\subsubsection{Interpretation.}

The interpretation of the Green's function is quite straightforward:
to propagate from one point to the other there are four kinds of basic
trajectories (see Figure 2), (1) one can go straight from the starting
point to the arrival point, or (2-3) go after bouncing against one
of the two boundaries, and (4) lastly go after bouncing two times
against different boundaries. These basic trajectories must then be
convoluted by round trips along the whole loop (of length $2L$) which
yield the overall factor $\left(1-r_{S}r_{D}e^{i2\varphi}\right)^{-1}$
(where $\varphi=\frac{\omega}{u}L$) in the frequency domain expression
in eq.(\ref{eq:green}).

The main difference for the chiral propagators is the appearance of
zero modes. The other terms have straightforward interpretations:
as before they correspond to straight trajectories from $y$ to $x$
or to propagation with bouncing at either or both of the boundaries.
The $\theta(x-y)$ or $\theta(y-x)$ come from the fact that chiral
propagation prevents some trajectories depending on the respective
positions of $x$ and $y$.

\begin{figure}
\includegraphics{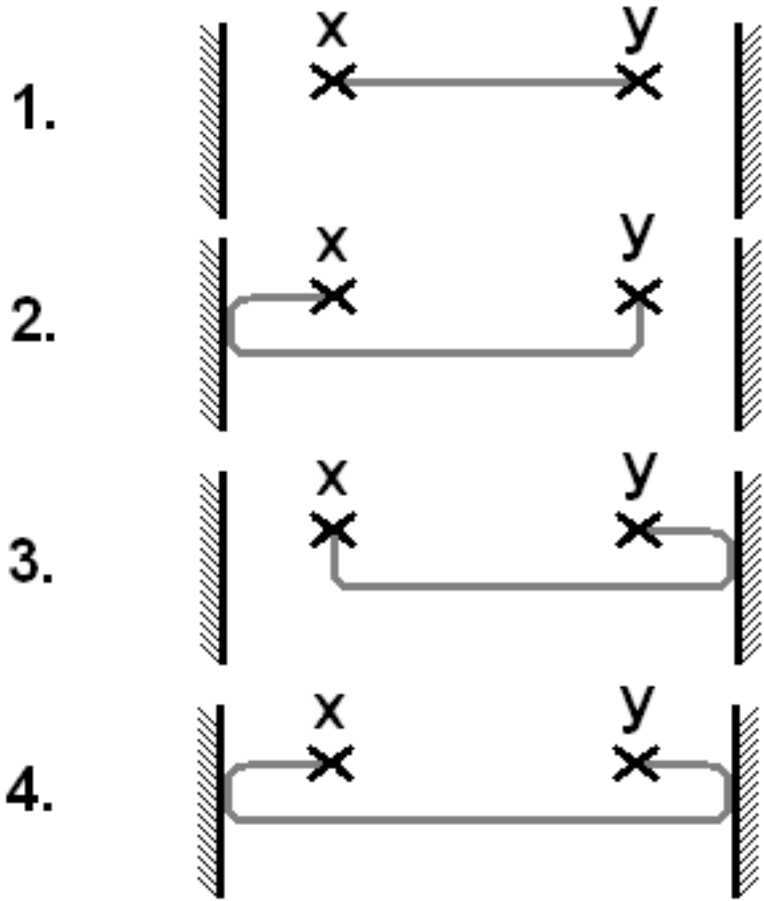}

\caption{Interpretation of the Green's function.}
\end{figure}

We have for now only discussed the retarded Green's function but of
course the advanced Green's function is simply related to the retarded
one through:

\[
G_{A}(x,t;\, y,0)=G_{R}(y,-t;\, x,0).\]

So there is no more work to do. The causal Green's function can also
be computed: for instance from the two point correlator (which is
discussed later). 

We also observe that the retarded and advanced Green's functions do
not depend on temperature since the equation of motion obeyed by the
Green's functions has no temperature dependence. We have therefore
found the phase field Green's functions \emph{at all temperatures}.

\subsubsection{Poles and relation to open boundary LL.}

The open boundary conditions (OBC) are a limiting case of the IBC
considered in this paper. Hard-walls at the boundaries can be reproduced
by considering infinite resistances $Z{}_{S}$ and $Z_{D}$ : this
then implies that the reflection coefficients are equal to unity $r_{S}=r_{D}=1$).
It can then be checked that our expression for the Green's function
reduces to that found for the LL with open boundary conditions. We
refer to Appendix B. Open boundary conditions are actually closely
related to the ones discussed in this paper: indeed the excitations
in the LL with IBC are those of the LL with OBC albeit with a finite-lifetime.
While in the infinite LL or the PBC LL one has travelling waves these
excitations are just the standing waves expected from a system enclosed
within boundaries.

Indeed the poles are simply:\begin{equation}
\omega_{n}=\frac{n\pi u}{L}+i\frac{u}{2L}\,\ln(r_{S}r_{D}).\end{equation}
For the OBC there are nodes at the boundaries; the resonances are
therefore such that the length of the wire is $L=n\lambda/2$ which
leads to $\omega_{n}=\frac{n\pi u}{L}$. Within the simple model of
boundary resistances (real positive impedances) the lifetime $\tau=2L/\left(u\:\ln(r_{S}r_{D})\right)$
is independent of the index mode: the level broadening is constant
for each standing wave plasma mode. But more complicated situations
can be considered: if we assume \textbf{frequency dependent complex
impedances} the reflection coefficients then acquire a frequency dependence.
This however does not affect the validity of the expressions in the
frequency domain just derived for the Green's function. However the
structure of the poles will not be quite as simple as that described
above, since the poles are now determined by:

\[
1-r_{S}(\omega)r_{D}(\omega)e^{i2L\omega/u}=0.\]

How can we probe these poles? One of the simplest way is through conductivity
or conductance measurements. These poles will show up as resonances.
Indeed as shown in Appendix A the conductivity is simply:\[
\sigma(x,y,\omega)=\frac{e}{\pi}\,\omega\, G_{R}(x,y,\omega).\]

The conductance (which is a matrix in this context of a gated wire
connected to two electrodes) was computed elsewhere \cite{ac}.

\subsubsection{Impedance matching.}

Let us recall the basic physics of transmission lines: an ideal transmission
line or $LC$-line (e.g. a coaxial cable) has an energy per unit length
$\mathcal{E}(x)=\frac{1}{2}\mathcal{L}\, j^{2}(x)+\frac{1}{2\mathcal{C}}\,\rho^{2}(x)$
where $\mathcal{L}$ and $\mathcal{C}$ are respectively an inductance
and a capacitance per unit length. For an infinite transmission line
the eigenmodes are traveling waves (plasmons) with velocity $u=1/\sqrt{\mathcal{LC}}$.
For a finite transmission line connected to two resistors at both
boundaries one observes reflections: in general any pulse injected
in the transmission line is reflected which leads to energy losses.
In order to minimize losses electrical engineers take advantage of
the phenomenon of 'impedance matching': if the resistors have identical
resistances equal to $Z_{0}=\sqrt{\mathcal{L}/\mathcal{C}}$ (the
characteristic impedance of the transmission line) then the reflection
coefficients vanish so that no reflections can occur in the combined
system of two resistors+transmission line, which becomes effectively
lossless. The finite transmission line has become effectively equivalent
to an infinite transmission line. This is the origin of the normalized
characteristic impedance of coaxial cables.

What is the relation to the Luttinger liquid ? A LL is actually a
quantum transmission line. Indeed its Hamitonian density is just that
of a quantum $LC$-line since: 

\[
H=\intop dx\:\frac{hu}{4K}\,\rho^{2}+\frac{huK}{4}\, j^{2}\]
which rewritten in terms of the charge density and the charge current
$\rho_{e}=e\:\rho$ and $j_{e}=euK\: j$ becomes:\[
H=\intop dx\:\frac{\mathcal{L}}{2}\: j_{e}^{2}+\frac{1}{2\mathcal{C}}\:\rho_{e}^{2}\]
with:\[
\mathcal{\mathcal{L}}=\frac{h}{2uKe^{2}},\;\mathcal{C}=\frac{2Ke^{2}}{hu}.\]

So it is only natural to inquire whether the 'impedance matching'
physics is still valid at the quantum level. Quite remarkably it is.
We prove the following theorem:

\vspace{1.5cm}

\textbf{Theorem:} the physics of a finite length Luttinger liquid
connected to two resistors having resistances equal to the characteristic
impedance $Z_{S}=Z_{D}=Z_{0}$ is equivalent to that of an infinite
LL (for any observable defined on the length of the LL).

\textbf{Proof:} the proof follows from identity of the Green's functions.
Indeed \textbf{}$Z_{0}=Z_{S}=Z_{D}$ implies: $r_{S}=r_{D}=0$. The
expressions of the one-body Green's functions we have computed then
trivially reduce to those of an infinite LL. The N-body Green's functions
are therefore also equal since by Wick's theorem they reduce to a
product of single particle Green's functions.\vspace{1.5cm}

\begin{figure}
\includegraphics{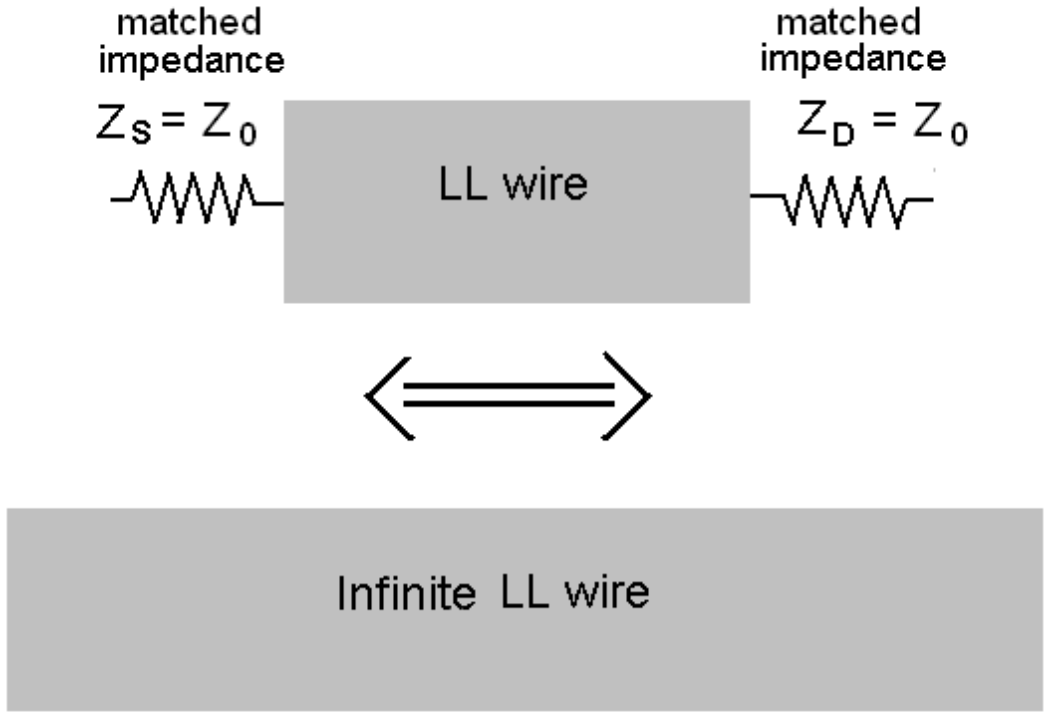}

\caption{Impedance matching.}
\end{figure}

Here in the quantum case the cancellation of the reflection coefficients
leads to the same physics as in the classical case. Note that the
result is still valid if we assume frequency dependent reflection
coefficients: if one checks carefully our derivation of the retarded
Green's function in section \ref{sec:Green's-function-of} one will
notice that the expressions do not require frequency-independent coefficients. 

Such a result might be useful in any situation where the intrinsic
properties of a LL (infinite system) are needed: for transport experiments
on quantum wires or carbon nanotubes the coupling to the leads (source
and drain) unavoidably modifies the pure physics of the LL. Should
one be able to tune the impedance of the leads, one might be able
to disenfranchise oneself from the interfering effects of the leads.

In general however the leads will couple to the LL not only with a
resistive component but also a capacitive (or even an inductive) component.
To achieve perfect impedance matching would mean to be able for all
frequency to adjust $Z_{S}(\omega)=Z_{0}=Z_{D}(\omega)$. For all
practical purposes depending on the phenomenon one wishes to observe
it will be enough to match impedances on a finite window.

Note also that the possibility to match impedances rests crucially
on an independent measurement of the characteristic impedance $Z{}_{0}$.
This can be achieved through a variety of means: for instance through
the finite-frequency conductance as explained by the author in \cite{ac}
or through tunneling experiments. 

However it might not be completely necessary to measure the characteristic
impedance: one might think of time-resolved experiments where one
sends a charged pulse in the wire; if one detects reflected pulses
this means that impedances are not matched (actually such tests are
routinely used by telecom operators on their transmission lines to
find broken lines).

\section{Two-point Correlators of the phase fields.\label{sec:Two-point-Correlators-of}}

We derive in this sections several correlators at zero and finite-temperature.
They will be needed for the proof of bosonization but are also interesting
in themselves since they are the building blocks for the vertex operators
correlators. All observables of interest can then be calculated: tunneling
density of states, current noise, etc.

\subsection{Correlators of the fields $\phi$ and $\Theta$.}

It is a simple matter to extract two-point correlation functions;
using the definitions of the Green's functions one has indeed at zero
temperature: \[
\left\langle \phi(x,\omega)\phi(y,-\omega)\right\rangle =i\,\theta(\omega)\:\left[G_{R}(x,y,\omega)-G_{A}(x,y,\omega)\right].\]

Tedious but uneventful calculations finally yield the following correlator
at zero temperature :\begin{eqnarray}
F(x,y,t)= & \left\langle \phi(x,t)\phi(y,0)\right\rangle \label{eq:phicorrel}\\
= & -\frac{K}{4\pi}\;\sum_{n=-\infty}^{+\infty}\left(r_{S}r_{D}\right)^{\left|n\right|} & \,\ln\left\{ \left[\delta+i(t+\frac{2nL}{u})\right]^{2}+(\frac{x-y}{u})^{2}\right\} \nonumber \\
+ & \frac{K}{4\pi}\;\sum_{n=0}^{+\infty}\left(r_{S}r_{D}\right)^{n}\: r_{S} & \,\ln\prod_{\varepsilon=\pm1}\left[\delta+i(t+\varepsilon\frac{x+y+(2n+1)L}{u})\right]\nonumber \\
+ & \frac{K}{4\pi}\;\sum_{n=0}^{+\infty}\left(r_{S}r_{D}\right)^{n}\: r_{D} & \,\ln\prod_{\varepsilon=\pm1}\left[\delta+i(t+\varepsilon\frac{x+y-(2n+1)L}{u})\right].\nonumber \end{eqnarray}

Extension to finite-temperature is done by observing that the retarded
and advanced Green's function are temperature independent. Let us
consider then the commutator:\begin{eqnarray*}
M(x,t;y,0) & = & \left\langle \left[\phi(x,t);\,\phi(y,0)\right]\right\rangle .\\
 & = & \left\langle \phi(x,t)\,\phi(y,0)-\phi(y,0)\,\phi(x,t)\right\rangle \end{eqnarray*}
By fluctuation-dissipation (Lehmann's spectral decomposition):\begin{eqnarray*}
M(x,y,\omega) & = & (1-e^{-\beta\omega})\left\langle \phi(x,\omega)\,\phi(y,-\omega)\right\rangle _{T}\\
 & = & (e^{\beta\omega}-1)\left\langle \phi(y,-\omega)\,\phi(x,\omega)\right\rangle _{T}\end{eqnarray*}
at temperature $T$. At zero temperature this implies:\[
M(x,y,\omega)=\theta(\omega)\left\langle \phi(x,\omega)\,\phi(y,-\omega)\right\rangle _{0}-\theta(-\omega)\left\langle \phi(y,-\omega)\,\phi(x,\omega)\right\rangle _{0}\]
where the subscript denotes zero temperature. Since $M$ is actually
temperature independent, comparison of the two expressions yields:\[
F(x,y,\omega,T)=\left\langle \phi(x,\omega)\,\phi(y,-\omega)\right\rangle _{T}=\frac{1}{1-e^{-\beta\omega}}\left[\theta(\omega)\, F(x,y,\omega,T=0)-\theta(-\omega)\, F(y,x,\omega,T=0)\right]\]
so that the finite-temperature correction is:\[
F(x,y,\omega,T)-F(x,y,\omega,T=0)=\theta(\omega)\,\frac{1}{e^{\beta\omega}-1}\, F(x,y,\omega,T=0)-\theta(-\omega)\,\frac{1}{1-e^{-\beta\omega}}\, F(y,x,\omega,T=0).\]

To regulate the previous correlator one defines: $\widehat{F}_{-+}(x,y,t)=\left\langle \phi(x,t)\,\phi(y,0)\right\rangle -\frac{1}{2}\left\langle \phi^{2}(x,t)+\phi^{2}(y,0)\right\rangle $.

The finite-temperature correlator acquires then a correction from
its zero temperature expression $\Delta F=\widehat{F}_{-+}(x,y,t,T)-\widehat{F}_{-+}(x,y,t,T=0)$:\begin{eqnarray*}
\Delta F=-\frac{K}{4\pi}\;\sum_{n=-\infty}^{+\infty}\left(r_{S}r_{D}\right)^{\left|n\right|}\,\ln\left\{ \frac{\left|\Gamma\left[1+i\beta^{-1}2nL)\right]\right|^{4}}{\left|\Gamma\left[1+i\beta^{-1}(t+\frac{x-y+2nL}{u})\right]\Gamma\left[1+i\beta^{-1}(t+\frac{y-x+2nL}{u})\right]\right|^{2}}\right\} \\
+\frac{K}{4\pi}\;\sum_{n=0}^{+\infty}\left(r_{S}r_{D}\right)^{n}\: r_{S}\:\ln\left\{ \frac{\left|\Gamma\left[1+i\beta^{-1}(\frac{2x+(2n+1)L}{u})\right]\right|^{2}\left|\Gamma\left[1+i\beta^{-1}(\frac{2y+(2n+1)L}{u})\right]\right|^{2}}{\left|\Gamma\left[1+i\beta^{-1}(t+\frac{x+y+(2n+1)L}{u})\right]\right|^{2}\left|\Gamma\left[1+i\beta^{-1}(t-\frac{x+y+(2n+1)L}{u})\right]\right|^{2}}\right\} \\
+\frac{K}{4\pi}\;\sum_{n=0}^{+\infty}\left(r_{S}r_{D}\right)^{n}\: r_{D}\:\ln\left\{ \frac{\left|\Gamma\left[1+i\beta^{-1}(\frac{2x-(2n+1)L}{u})\right]\right|^{2}\left|\Gamma\left[1+i\beta^{-1}(\frac{2y-(2n+1)L}{u})\right]\right|^{2}}{\left|\Gamma\left[1+i\beta^{-1}(t+\frac{x+y-(2n+1)L}{u})\right]\right|^{2}\left|\Gamma\left[1+i\beta^{-1}(t-\frac{x+y-(2n+1)L}{u})\right]\right|^{2}}\right\} \end{eqnarray*}
where we have made use of the product expansion of the Gamma function
$\Gamma$ and $\beta$ is $\frac{1}{T}$. This can also be rexpressed
as:\begin{eqnarray*}
\Delta F & = & -\frac{K}{4\pi}\;\sum_{n=-\infty}^{+\infty}\left(r_{S}r_{D}\right)^{\left|n\right|}\,\ln\left\{ \frac{\left[sh_{c}\pi\beta^{-1}(t+\frac{x-y+2nL}{u})\right]\:\left[sh_{c}\pi\beta^{-1}(t+\frac{y-x+2nL}{u})\right]}{\left[sh_{c}\pi\beta^{-1}2nL\right]^{2}}\right\} \\
 & + & \frac{K}{4\pi}\;\sum_{n=0}^{+\infty}\left(r_{S}r_{D}\right)^{n}\: r_{S}\:\ln\left\{ \frac{\left[sh_{c}\pi\beta^{-1}(t+\frac{x+y+(2n+1)L}{u})\right]\:\left[sh_{c}\pi\beta^{-1}(t-\frac{x+y+(2n+1)L}{u})\right]}{\left[sh_{c}\pi\beta^{-1}(\frac{2x+(2n+1)L}{u})\right]\left[sh_{c}\pi\beta^{-1}(\frac{2y+(2n+1)L}{u})\right]}\right\} \\
 & + & \frac{K}{4\pi}\;\sum_{n=0}^{+\infty}\left(r_{S}r_{D}\right)^{n}\: r_{D}\:\ln\left\{ \frac{\left[sh_{c}\pi\beta^{-1}(t+\frac{x+y-(2n+1)L}{u})\right]\:\left[sh_{c}\pi\beta^{-1}(t-\frac{x+y-(2n+1)L}{u})\right]}{\left[sh_{c}\pi\beta^{-1}(\frac{2x-(2n+1)L}{u})\right]\left[sh_{c}\pi\beta^{-1}(\frac{2y-(2n+1)L}{u})\right]}\right\} \end{eqnarray*}
whare we have defined the function $sh_{c}x=\sinh x/x$.

The correlator for the other phase field $\Theta$ can be found from
the previous expressions by the operations: $(r_{S},r_{D},K)\rightarrow(-r_{S},-r_{D},1/K)$.
The change of sign for the reflection coefficients follows from the
fact that the chiral components of $\Theta$ are $\Theta_{\pm}=\pm\phi_{\pm}/K$
so that the reflection coefficients for $\Theta$ acquire a relative
minus sign with respect to those for $\phi$.

\subsection{Chiral correlators.}

The chiral correlators will be useful for computing the cross-correlators
of the phase fields. We use again:\begin{eqnarray*}
\left\langle \phi_{\pm}(x,\omega)\phi_{\pm}(y,-\omega)\right\rangle  & = & i\,\theta(\omega)\:\left[G_{R}^{\pm\pm}(x,y,\omega)-G_{A}^{\pm\pm}(x,y,\omega)\right].\\
 & (= & -2\,\theta(\omega)\:\Im mG_{R}^{\pm\pm}(x,y,\omega)\:).\end{eqnarray*}

Using the expressions of the chiral retarded propagators and since
$G_{A}^{\pm\pm}(x,y,\omega)=G_{R}^{\pm\pm}(y,x,-\omega)$ and $G_{A}^{\pm\mp}(x,y,\omega)=G_{R}^{\mp\pm}(y,x,-\omega)$
one finds finally:\begin{eqnarray*}
\left\langle \phi_{\pm}(x,t)\phi_{\pm}(y,0)\right\rangle =-\frac{K}{4\pi}\sum_{n=-\infty}^{+\infty}\left(r_{S}r_{D}\right)^{\left|n\right|}\ln\left[\delta+i\left(t-\frac{2nL\pm(x-y)}{u}\right)\right]\\
\left\langle \phi_{+}(x,t)\phi_{-}(y,0)\right\rangle =\frac{K}{4\pi}\sum_{n=0}^{+\infty}\left(r_{S}r_{D}\right)^{n}\:\left\{ r_{S}\ln\left[\delta+i\left(t-\frac{x+y+(2n+1)L}{u}\right)\right]\right.\\
\left.+r_{D}\ln\left[\delta+i\left(t-\frac{x+y-(2n+1)L}{u}\right)\right]\right\} -i\frac{K}{8}+i\frac{K}{8}\:\frac{r_{S}-r_{D}}{1-r_{S}r_{D}}\\
\left\langle \phi_{-}(x,t)\phi_{+}(y,0)\right\rangle =\frac{K}{4\pi}\sum_{n=0}^{+\infty}\left(r_{S}r_{D}\right)^{n}\:\left\{ r_{S}\ln\left[\delta+i\left(t+\frac{x+y+(2n+1)L}{u}\right)\right]\right.\\
\left.+r_{D}\ln\left[\delta+i\left(t+\frac{x+y-(2n+1)L}{u}\right)\right]\right\} +i\frac{K}{8}-i\frac{K}{8}\:\frac{r_{S}-r_{D}}{1-r_{S}r_{D}}\end{eqnarray*}

Note the presence of non-trivial zero-mode terms for the cross-correlators
of the chiral fields.

For later use in the bosonization proof it will be useful to consider
these expressions for $K=1$. To avoid confusion we define the chiral
fields for $K=1$ as $\varphi_{R}$ and $\varphi_{L}$; their correlators
are therefore as above with $K=1$:\begin{eqnarray*}
\left\langle \varphi_{R/L}(x,t)\varphi_{R/L}(y,0)\right\rangle =-\frac{1}{4\pi}\sum_{n=-\infty}^{+\infty}\left(r_{S}r_{D}\right)^{\left|n\right|}\ln\left[\delta+i\left(t-\frac{2nL\pm(x-y)}{u}\right)\right];\\
\left\langle \varphi_{R}(x,t)\varphi_{L}(y,0)\right\rangle =\frac{1}{4\pi}\sum_{n=0}^{+\infty}\left(r_{S}r_{D}\right)^{n}\:\left\{ r_{S}\ln\left[\delta+i\left(t-\frac{x+y+(2n+1)L}{u}\right)\right]\right.\\
\left.+r_{D}\ln\left[\delta+i\left(t-\frac{x+y-(2n+1)L}{u}\right)\right]\right\} -\frac{i}{8}+\frac{i}{8}\:\frac{r_{S}-r_{D}}{1-r_{S}r_{D}};\\
\left\langle \varphi_{L}(x,t)\varphi_{R}(y,0)\right\rangle =\frac{1}{4\pi}\sum_{n=0}^{+\infty}\left(r_{S}r_{D}\right)^{n}\:\left\{ r_{S}\ln\left[\delta+i\left(t+\frac{x+y+(2n+1)L}{u}\right)\right]\right.\\
\left.+r_{D}\ln\left[\delta+i\left(t+\frac{x+y-(2n+1)L}{u}\right)\right]\right\} +\frac{i}{8}-\frac{i}{8}\:\frac{r_{S}-r_{D}}{1-r_{S}r_{D}}.\end{eqnarray*}

\subsection{Cross correlators of $\phi$ and $\Theta$.}

Since:\begin{eqnarray*}
\phi & = & \phi_{+}+\phi_{-}\\
\Theta & = & \frac{\phi_{+}-\phi_{-}}{K}\end{eqnarray*}
one has: \begin{eqnarray*}
\left\langle \phi\:\Theta\right\rangle  & =K^{-1} & \left\{ \left\langle \phi_{+}\phi_{+}\right\rangle -\left\langle \phi_{-}\phi_{-}\right\rangle -\left\langle \phi_{+}\phi_{-}\right\rangle +\left\langle \phi_{-}\phi_{+}\right\rangle \right\} \\
\left\langle \Theta\:\phi\right\rangle  & =K^{-1} & \left\{ \left\langle \phi_{+}\phi_{+}\right\rangle -\left\langle \phi_{-}\phi_{-}\right\rangle +\left\langle \phi_{+}\phi_{-}\right\rangle -\left\langle \phi_{-}\phi_{+}\right\rangle \right\} \end{eqnarray*}
Finally:\begin{eqnarray*}
\left\langle \phi(x,t)\Theta(y,0)\right\rangle  & = & -\frac{1}{4\pi}\;\sum_{n=-\infty}^{+\infty}\left(r_{S}r_{D}\right)^{\left|n\right|}\ln\left\{ \frac{\delta+i\left(t+\frac{2nL+y-x}{u}\right)}{\delta+i\left(t+\frac{2nL+x-y}{u}\right)}\right\} \\
 &  & -\frac{1}{4\pi}\;\sum_{n=0}^{+\infty}\left(r_{S}r_{D}\right)^{n}\: r_{S}\ln\left\{ \frac{\delta+i\left(t-\frac{x+y+(2n+1)L}{u}\right)}{\delta+i\left(t+\frac{x+y+(2n+1)L}{u}\right)}\right\} \\
 &  & -\frac{1}{4\pi}\;\sum_{n=0}^{+\infty}\left(r_{S}r_{D}\right)^{n}\: r_{D}\ln\left\{ \frac{\delta+i\left(t-\frac{x+y-(2n+1)L}{u}\right)}{\delta+i\left(t+\frac{x+y-(2n+1)L}{u}\right)}\right\} \\
 &  & +\frac{i}{4}-\frac{i}{4}\:\frac{r_{S}-r_{D}}{1-r_{S}r_{D}}\end{eqnarray*}
\begin{eqnarray*}
\left\langle \Theta(x,t)\phi(y,0)\right\rangle  & = & -\frac{1}{4\pi}\;\sum_{n=-\infty}^{+\infty}\left(r_{S}r_{D}\right)^{\left|n\right|}\ln\left\{ \frac{\delta+i\left(t+\frac{2nL+y-x}{u}\right)}{\delta+i\left(t+\frac{2nL+x-y}{u}\right)}\right\} \\
 &  & +\frac{1}{4\pi}\;\sum_{n=0}^{+\infty}\left(r_{S}r_{D}\right)^{n}\: r_{S}\ln\left\{ \frac{\delta+i\left(t-\frac{x+y+(2n+1)L}{u}\right)}{\delta+i\left(t+\frac{x+y+(2n+1)L}{u}\right)}\right\} \\
 &  & +\frac{1}{4\pi}\;\sum_{n=0}^{+\infty}\left(r_{S}r_{D}\right)^{n}\: r_{D}\ln\left\{ \frac{\delta+i\left(t-\frac{x+y-(2n+1)L}{u}\right)}{\delta+i\left(t+\frac{x+y-(2n+1)L}{u}\right)}\right\} \\
 &  & -\frac{i}{4}+\frac{i}{4}\:\frac{r_{S}-r_{D}}{1-r_{S}r_{D}}\end{eqnarray*}
At finite temperature the regularized correlators $F_{1}=\left\langle \phi(x,t)\Theta(y,0)\right\rangle -\frac{1}{2}\left\langle \phi(x,0)\Theta(x,0)\right\rangle -\frac{1}{2}\left\langle \phi(y,0)\Theta(y,0)\right\rangle $
and $F_{2}=\left\langle \Theta(x,t)\phi(y,0)\right\rangle -\frac{1}{2}\left\langle \Theta(x,0)\phi(x,0)\right\rangle -\frac{1}{2}\left\langle \Theta(y,0)\phi(y,0)\right\rangle $
get the corrections:\begin{eqnarray*}
\Delta F_{1} & = & -\frac{1}{4\pi}\;\sum_{n=-\infty}^{+\infty}\left(r_{S}r_{D}\right)^{\left|n\right|}\ln\left\{ \frac{sh_{c}\pi\beta^{-1}\left(t+\frac{2nL+y-x}{u}\right)}{sh_{c}\pi\beta^{-1}\left(t+\frac{2nL+x-y}{u}\right)}\right\} \\
 &  & -\frac{1}{4\pi}\;\sum_{n=0}^{+\infty}\left(r_{S}r_{D}\right)^{n}\: r_{S}\ln\left\{ \frac{sh_{c}\pi\beta^{-1}\left(t-\frac{x+y+(2n+1)L}{u}\right)}{sh_{c}\pi\beta^{-1}\left(t+\frac{x+y+(2n+1)L}{u}\right)}\right\} \\
 &  & -\frac{1}{4\pi}\;\sum_{n=0}^{+\infty}\left(r_{S}r_{D}\right)^{n}\: r_{D}\ln\left\{ \frac{sh_{c}\pi\beta^{-1}\left(t-\frac{x+y-(2n+1)L}{u}\right)}{sh_{c}\pi\beta^{-1}\left(t+\frac{x+y-(2n+1)L}{u}\right)}\right\} \end{eqnarray*}
and:\begin{eqnarray*}
\Delta F_{2} & = & -\frac{1}{4\pi}\;\sum_{n=-\infty}^{+\infty}\left(r_{S}r_{D}\right)^{\left|n\right|}\ln\left\{ \frac{sh_{c}\pi\beta^{-1}\left(t+\frac{2nL+y-x}{u}\right)}{sh_{c}\pi\beta^{-1}\left(t+\frac{2nL+x-y}{u}\right)}\right\} \\
 &  & +\frac{1}{4\pi}\;\sum_{n=0}^{+\infty}\left(r_{S}r_{D}\right)^{n}\: r_{S}\ln\left\{ \frac{sh_{c}\pi\beta^{-1}\left(t-\frac{x+y+(2n+1)L}{u}\right)}{sh_{c}\pi\beta^{-1}\left(t+\frac{x+y+(2n+1)L}{u}\right)}\right\} \\
 &  & +\frac{1}{4\pi}\;\sum_{n=0}^{+\infty}\left(r_{S}r_{D}\right)^{n}\: r_{D}\ln\left\{ \frac{sh_{c}\pi\beta^{-1}\left(t-\frac{x+y-(2n+1)L}{u}\right)}{sh_{c}\pi\beta^{-1}\left(t+\frac{x+y-(2n+1)L}{u}\right)}\right\} \end{eqnarray*}
where as before we have defined $sh_{c}(x)=\sinh x/x.$

\section{Bosonization.\label{sec:Bosonization.}}

It is far from obvious that bosonization works with the boundary conditions
considered in this paper.

We now show it does or more precisely we show that:

\emph{The bosonic field theory on a finite length with boundary conditions}
\begin{eqnarray*}
H & = & \int_{-L/2}^{L/2}dx\:\left(\partial_{x}\varphi_{R}\right)^{2}+\left(\partial_{x}\varphi_{L}\right)^{2}\\
\partial_{x}\varphi_{R}(-L/2) & = & r_{S}\:\partial_{x}\varphi_{L}(-L/2)\\
\partial_{x}\varphi_{L}(L/2) & = & r_{D}\,\partial_{x}\varphi_{R}(L/2)\end{eqnarray*}
\emph{with :} \begin{eqnarray*}
\left[\varphi_{R/L}(x,t),\varphi_{R/L}(y,t)\right] & = & \pm\frac{i}{4}sgn(x-y)\\
\left[\varphi_{R}(x,t),\varphi_{L}(y,t)\right] & =- & \frac{i}{4}\end{eqnarray*}
\emph{is equivalent to the fermionic field theory:}\begin{eqnarray*}
H & = & \int_{-L/2}^{L/2}dx\:-iR^{+}\partial_{x}R+iL^{+}\partial_{x}L\\
\rho_{R}(-L/2) & = & r_{S}\:\rho_{L}(-L/2)\\
\rho_{L}(L/2) & = & r_{D}\:\rho_{R}(L/2)\end{eqnarray*}
\emph{where} $R$ \emph{and} $L$ \emph{are Fermi fields obeying the
usual anticommutation rules and:} \begin{eqnarray*}
\rho_{R} & (x,t)= & :R^{+}(x,t)R(x,t):\\
\rho_{L} & (x,t)= & :L^{+}(x,t)L(x,t):\end{eqnarray*}
\emph{provided we identify:}\begin{eqnarray*}
R(x,t) & = & \frac{1}{\sqrt{2\pi\delta}}\, exp\,2i\sqrt{\pi}\varphi_{R}\\
L(x,t) & = & \frac{1}{\sqrt{2\pi\delta}}\, exp\,-2i\sqrt{\pi}\varphi_{L}\end{eqnarray*}
\emph{where} $\delta$ \emph{is a short-distance cut-off.}

Note that we have normalized the Fermi velocity to $u=1$. 

We proceed in three steps:

(1) we check that the vertex operators obey fermionic commutation
relations;

(2) by point-splitting we show that the usual relations still hold:
\begin{eqnarray*}
:\rho_{R} & (x,t):= & \frac{1}{\sqrt{\pi}}\partial_{x}\varphi_{R}\\
:\rho_{L} & (x,t):= & \frac{1}{\sqrt{\pi}}\partial_{x}\varphi_{L}\end{eqnarray*}

(3) and also by point splitting that: \begin{eqnarray*}
::\rho_{R} & (x,t):^{2}:=\frac{-i}{\pi} & :R^{+}\partial_{x}R:\\
::\rho_{L} & (x,t):^{2}:=\frac{i}{\pi} & :L^{+}\partial_{x}L:\end{eqnarray*}
from which the proof is trivially completed by substitution.

\textbf{Step 1:}

For $R(x,t)=\frac{1}{\sqrt{2\pi\delta}}exp\,2i\sqrt{\pi}\varphi_{R}$
the relation $\left\{ R(x,t);R(y,t)\right\} =0$ is automatically
fulfilled given the commutation relation of the chiral field.

Indeed using the identity :\[
e^{A}e^{B}=:e^{A+B}:\: e^{\left\langle AB+\frac{A^{2}+B^{2}}{2}\right\rangle }\]
it follows that \begin{eqnarray*}
R(x,t)R(y,t) & = & R(y,t)R(x,t)\: e^{-4\pi\left\langle \left[\varphi_{R}(x,t),\varphi_{R}(y,t)\right]\right\rangle }\\
 & = & R(y,t)R(x,t)\: e^{-i\pi\, sgn(x-y)}\\
 & = & -R(y,t)R(x,t)\end{eqnarray*}
Similarly $\left\{ R(x,t);L(y,t)\right\} =0$ using $\left[\varphi_{R}(x,t),\varphi_{L}(y,t)\right]=-i/4$.

This step is trivial but it remains to show that $\left\{ R(x,t),R^{+}(y,t)\right\} =\delta(x-y)$
which is less straightforward. 

We use the expressions derived for the chiral correlators (inserting
$K=1$ and $u=1$) (note that to make clear that the fields are taken
at $K=1$ we use the notations $\varphi_{R}$ and $\varphi_{L}$ instead
of $\phi_{\pm}$ in the whole section):\[
\left\langle \varphi_{R}(x,t)\varphi_{R}(y,0)\right\rangle =-\frac{1}{4\pi}\sum_{n=-\infty}^{+\infty}\left(r_{S}r_{D}\right)^{\left|n\right|}\ln\left[\delta+i\left(t-2nL\pm(x-y)\right)\right].\]

This entails: 

\noindent \begin{flushleft}
\begin{eqnarray*}
 & \left\{ R(x,t),R^{+}(y,t)\right\} =\frac{1}{2\pi\delta}\::exp\,2i\sqrt{\pi}\left(\varphi_{R}(x,t)-\varphi_{R}(y,t)\right):\\
 & \times\left\{ \exp\,4\pi\left\langle \varphi_{R}(x,t)\varphi_{R}(y,t)-\frac{1}{2}\left(\varphi_{R}^{2}(x,t)+\varphi_{R}^{2}(y,t)\right)\right\rangle +\right.\\
 & \left.+\exp\,4\pi\left\langle \varphi_{R}(y,t)\varphi_{R}(x,t)-\frac{1}{2}\left(\varphi_{R}^{2}(x,t)+\varphi_{R}^{2}(y,t)\right)\right\rangle \right\} \\
= & \frac{1}{2\pi\delta}\::exp\,2i\sqrt{\pi}\left(\varphi_{R}(x,t)-\varphi_{R}(y,t)\right):\\
 & \times\left\{ \exp\,-\sum_{n=-\infty}^{+\infty}\left(r_{S}r_{D}\right)^{\left|n\right|}\:\ln\left(\frac{\delta+i(2nL+x-y)}{\delta+i2nL}\right)+\right.\\
 & \left.+\exp\,-\sum_{n=-\infty}^{+\infty}\left(r_{S}r_{D}\right)^{\left|n\right|}\:\ln\left(\frac{u\delta+i(2nL+y-x)}{u\delta+i2nL}\right)\right\} \\
= & \frac{1}{2\pi\delta}\::exp\,2i\sqrt{\pi}\left(\varphi_{R}(x,t)-\varphi_{R}(y,t)\right):\\
 & \times\left\{ \exp\,-\ln(\frac{\delta-i(x-y)}{\delta})\:\exp\,-\sum_{n\neq0}\left(r_{S}r_{D}\right)^{\left|n\right|}\:\ln\left(\frac{2nL+y-x}{2nL}\right)\right.\\
 & \left.+\exp\,-\ln(\frac{\delta-i(y-x)}{\delta})\:\exp\,-\sum_{n\neq0}\left(r_{S}r_{D}\right)^{\left|n\right|}\:\ln\left(\frac{2nL+x-y}{2nL}\right)\right\} \\
= & \frac{1}{2\pi}:exp\,2i\sqrt{\pi}\left(\varphi_{R}(x,t)-\varphi_{R}(y,t)\right):\exp\,-\sum_{n\neq0}\left(r_{S}r_{D}\right)^{n}\:\ln\left(\frac{2nL+y-x}{2nL}\right)\\
 & \times\left\{ \frac{1}{\delta-i(x-y)}+\frac{1}{\delta+i(x-y)}\right\} \\
= & :exp\,2i\sqrt{\pi}\left(\varphi_{R}(x,t)-\varphi_{R}(y,t)\right):\:\exp\,-\sum_{n\neq0}\left(r_{S}r_{D}\right)^{n}\:\ln\left(\frac{2nL+y-x}{2nL}\right)\:\delta(x-y)\\
= & \delta(x-y).\end{eqnarray*}

\par\end{flushleft}

\textbf{Step two:}

\begin{eqnarray*}
R^{+}(x)R(x) & = & \lim_{\varepsilon\rightarrow0}\frac{1}{2\pi\delta}\:\exp\,-2i\sqrt{\pi}\varphi_{R}(x+\varepsilon)\: exp\,2i\sqrt{\pi}\varphi_{R}(x)\\
 & = & \lim_{\varepsilon\rightarrow0}\frac{1}{2\pi\delta}\::\exp\,-2i\sqrt{\pi}\left(\varphi_{R}(x+\varepsilon)-\varphi_{R}(x)\right):\:\\
 &  & \exp\,4\pi\left\langle \varphi_{R}(x+\varepsilon)\varphi_{R}(x)-\frac{1}{2}\left(\varphi_{R}^{2}(x+\varepsilon)+\varphi_{R}^{2}(x)\right)\right\rangle \\
 & = & \lim_{\varepsilon\rightarrow0}\frac{1}{2\pi\delta}\:\left[1-2i\sqrt{\pi}\varepsilon\partial_{x}\varphi_{R}(x)\right]\:\exp\,-\sum_{n=-\infty}^{+\infty}\left(r_{S}r_{D}\right)^{\left|n\right|}\:\ln\left(\frac{\delta+i(2nL+\varepsilon)}{\delta+i2nL}\right)\\
 & = & \lim_{\varepsilon\rightarrow0}\frac{1}{2\pi\delta}\:\left[1-2i\sqrt{\pi}\varepsilon\partial_{x}\varphi_{R}(x)\right]\:\frac{\delta}{\delta-i\varepsilon}\:\exp\,-\sum_{n\neq0}\left(r_{S}r_{D}\right)^{\left|n\right|}\:\ln\left(\frac{i(2nL+\varepsilon)}{i2nL}\right)\end{eqnarray*}
In the last line the regulator can safely be put to zero in the exponential
(except for the $n=0$ term) since the $2nL$ term is finite. $\delta$
being the short-distance cut-off (the inverse of the bandwidth) is
smaller than any distance, so smaller than $\varepsilon$. The limit
$\delta\rightarrow0$ is therefore taken before the $\varepsilon\rightarrow0$
:

\[
R^{+}(x)R(x)=\lim_{\varepsilon\rightarrow0}\frac{1}{2\pi(-i\varepsilon)}\:\left[1-2i\sqrt{\pi}\varepsilon\partial_{x}\varphi_{R}(x)\right]\,\exp\,-\sum_{n\neq0}\left(r_{S}r_{D}\right)^{\left|n\right|}\:\ln\left(\frac{i(2nL+\varepsilon)}{i2nL}\right)\]
and we get after normal ordering (discarding a c-number piece) the
expected result. So: \begin{eqnarray*}
\rho_{R} & (x,t)= & :R^{+}(x,t)R(x,t):=\frac{1}{\sqrt{\pi}}\partial_{x}\varphi_{R}\\
\rho_{L} & (x,t)= & :L^{+}(x,t)L(x,t):=\frac{1}{\sqrt{\pi}}\partial_{x}\varphi_{L}\end{eqnarray*}

\textbf{Step three:}

We will need the correlators of the vertex operators.\begin{eqnarray*}
\left\langle R^{+}(x)R(y)\right\rangle  & = & \frac{1}{2\pi\delta}\:\left\langle :\exp\,-2i\sqrt{\pi}\left(\varphi_{R}(x)-\varphi_{R}(y)\right):\right\rangle \:\exp\,4\pi\left\langle \varphi_{R}(x)\varphi_{R}(y)-\frac{1}{2}\left(\varphi_{R}^{2}(x)+\varphi_{R}^{2}(y)\right)\right\rangle \\
 & = & \frac{1}{2\pi\delta}\:\exp\,-\sum_{n=-\infty}^{+\infty}\left(r_{S}r_{D}\right)^{\left|n\right|}\:\ln\left(\frac{\delta+i(2nL+x-y)}{\delta+i2nL}\right)\end{eqnarray*}
and:\begin{eqnarray*}
\left\langle R(x)R^{+}(y)\right\rangle  & = & \frac{1}{2\pi\delta}\:\left\langle :\exp\,2i\sqrt{\pi}\left(\varphi_{R}(x)-\varphi_{R}(y)\right):\right\rangle \:\exp\,4\pi\left\langle \varphi_{R}(x)\varphi_{R}(y)-\frac{1}{2}\left(\varphi_{R}^{2}(x)+\varphi_{R}^{2}(y)\right)\right\rangle \\
 & = & \frac{1}{2\pi\delta}\:\exp\,-\sum_{n=-\infty}^{+\infty}\left(r_{S}r_{D}\right)^{\left|n\right|}\:\ln\left(\frac{\delta+i(2nL+x-y)}{\delta+i2nL}\right).\end{eqnarray*}

We now compute the square of the normal ordered chiral density. This
is of course a singular operator which needs itself to be normal ordered.
By point splitting and then Wick's theorem:\begin{eqnarray*}
:R^{+}(x)R(x):^{2} & = & \lim_{\varepsilon\rightarrow0}:R^{+}(x+\varepsilon)R(x+\varepsilon)::R^{+}(x)R(x):\\
 & = & \lim_{\varepsilon\rightarrow0}:R^{+}(x+\varepsilon)R(x+\varepsilon)R^{+}(x)R(x):+\left\langle R^{+}(x+\varepsilon)R(x)\right\rangle :R(x+\varepsilon)R^{+}(x):\\
 & + & R^{+}(x+\varepsilon)R(x):\left\langle R(x+\varepsilon)R^{+}(x)\right\rangle +\left\langle R^{+}(x+\varepsilon)R(x)\right\rangle \left\langle R(x+\varepsilon)R^{+}(x)\right\rangle \\
 & = & \lim_{\varepsilon\rightarrow0}\frac{1}{2\pi\delta}\:\exp\,-\sum_{n=-\infty}^{+\infty}\left(r_{S}r_{D}\right)^{\left|n\right|}\:\ln\left(\frac{\delta+i(2nL+\varepsilon)}{\delta+i2nL}\right)\:\\
 &  & \left\{ :R(x+\varepsilon)R^{+}(x):+R^{+}(x+\varepsilon)R(x):\right\} \\
 & + & c-number.\end{eqnarray*}
As before in the exponential only the $n=0$ is really singular in
the limit $\delta\rightarrow0$; it cancels with the prefactor so
that after normal ordering:

\begin{eqnarray*}
::R^{+}(x)R(x):^{2} & := & \lim_{\varepsilon\rightarrow0}\frac{1}{2\pi i\varepsilon}\:\exp\,-\sum_{n\neq0}\left(r_{S}r_{D}\right)^{\left|n\right|}\:\ln\left(\frac{2nL+\varepsilon}{2nL}\right)\:\\
 &  & \left\{ :R(x+\varepsilon)R^{+}(x):+R^{+}(x+\varepsilon)R(x):\right\} \\
 & = & \lim_{\varepsilon\rightarrow0}\frac{1}{2\pi i\varepsilon}\:\left\{ :R(x+\varepsilon)R^{+}(x):+R^{+}(x+\varepsilon)R(x):\right\} \\
 & = & \frac{-i}{\pi}:R^{+}\partial_{x}R:\end{eqnarray*}
\textbf{Finally:} using the relations proven in step two the Hamiltonian
can be rewritten in terms of the currents of the fermion vertex operators:\[
H=\pi\int_{-L/2}^{L/2}Ldx\::\rho_{R}^{2}+\rho_{L}^{2}:\]
Then using the relation derived in step three:\begin{eqnarray*}
:\rho_{R} & (x,t):^{2}=\frac{-i}{\pi} & :R^{+}\partial_{x}R:\\
:\rho_{L} & (x,t):^{2}=\frac{i}{\pi} & :L^{+}\partial_{x}L:\end{eqnarray*}
implies immediately:\begin{eqnarray*}
H & = & \int_{-L/2}^{L/2}dx\:-iR^{+}\partial_{x}R+iL^{+}\partial_{x}L:\end{eqnarray*}

It remains to prove that the boundary conditions for the boson theory
translate into the quoted boundary conditions for the free fermion
theory: but this has already been shown in section \ref{sec:model}
of the paper (see eq.(\ref{bound1},\ref{coef},\ref{eq:bound2})
). 

\textbf{Switching on interactions.} What about interactions? Having
shown the relation between the free boson and the Dirac fermions,
interactions can now be switched on for the fermions. But since we
have proven that the customary dictionary of correspondence still
holds (vertex operator, currents ) it is clear that for the LL with
boundary conditions the transcription of fermion interactions will
go as in the standard LL. For example if we add the interaction:\[
V=\int_{-L/2}^{L/2}dx\: g_{2}\,\rho_{R}\rho_{L}+g_{4}\,\left(\rho_{R}{}^{2}+\rho_{L}{}^{2}\right)\]
the full Hamiltonian can be rewritten in terms of the phase fields
as:\[
H=\frac{1}{2}\int_{-L/2}^{L/2}dx\:(1+g_{4}+2g_{2})(\partial_{x}\phi)^{2}+(1+g_{4}-2g_{2})\,\Pi^{2}\]
which takes the standard form:\[
H=\frac{u}{2}\int_{-L/2}^{L/2}dx\,\,\frac{1}{K}\:\left(\partial_{x}\phi\right)^{2}+K\:\Pi^{2}\]
with \begin{eqnarray*}
K & = & \sqrt{\frac{1+g_{4}-2g_{2}}{1+g_{4}+2g_{2}}},\\
u & = & \sqrt{\left(1+g_{4}\right)^{2}-4g_{2}^{2}}.\end{eqnarray*}
(we have normalized the Fermi velocity $v_{F}$ to $1$ and also $\hbar=1$
in all this section). 

If we look closely at the bosonization proof we see that what does
the trick is always the $n=0$ term in the sum (this term corresponds
to reflectionless propagation in the propagator): this is sensible
since it corresponds to what one would have without reflections as
in the infinite system. Remarkably this shows that the ultralocal
structure of the current algebra is unaffected by the dissipative
boundaries.

What about spin? Extension to spinful LL is done in the same manner
as with the usual LL. We consider a second copy of the LL with IBC
and add a spin index $\sigma$. Given the fact that the boson fields
$\varphi_{R\sigma}$ and $\varphi_{L\sigma}$ for different spins
commute, as usual it suffices to add Majorana fermions $\eta_{\sigma}$
to enforce equal time commutation relations for the fields for different
spins \cite{mycitation} so that:\[
R_{\sigma}=\eta_{\sigma}\frac{1}{\sqrt{2\pi\delta}}\, exp\,2i\sqrt{\pi}\varphi_{R\sigma}\]
 and \[
\left\{ R_{\sigma}(x,t),R_{\sigma'}^{+}(y,t)\right\} =\delta_{\sigma\sigma'}\:\delta(x-y)\]
if $\left\{ \eta_{\sigma},\eta_{\sigma'}\right\} =2\delta_{\sigma\sigma'}$
and $\eta_{\sigma}^{+}=\eta_{\sigma}$.

\section{Fermion correlators.\label{sec:Fermion-correlators.}}

The fermion operators are given by the usual relation with the phase
fields :\begin{eqnarray*}
R(x,t) & = & \frac{1}{\sqrt{2\pi\delta}}\, exp\, i\sqrt{\pi}\left(\phi+\Theta\right),\\
L(x,t) & = & \frac{1}{\sqrt{2\pi\delta}}\,\, exp\, i\sqrt{\pi}\left(-\phi+\Theta\right),\end{eqnarray*}
up to unessential phases corresponding to shifts of the chemical potential
(see Appendix B and the discussion regarding open boundary conditions).

The two-point correlators are given by:

\begin{eqnarray*}
\left\langle R(x,t)R^{+}(y,0)\right\rangle = & \frac{1}{2\pi\delta}\,\exp\pi\left\langle \phi(x,t)\phi(y,0)\right\rangle +\left\langle \Theta(x,t)\Theta(y,0)\right\rangle +\left\langle \phi(x,t)\Theta(y,0)\right\rangle +\\
 & +\left\langle \Theta(x,t)\phi(y,0)\right\rangle -\frac{1}{2}\left[\left\langle \phi^{2}(x,t)+\phi^{2}(y,0)\right\rangle +\left\langle \Theta^{2}(x,t)+\Theta^{2}(y,0)\right\rangle +\right.\\
 & \left.\left.+\left\langle \phi(x,t)\Theta(x,t)+\phi(y,0)\Theta(y,0)\right\rangle +\left\langle \Theta(x,t)\phi(x,t)+\Theta(y,0)\phi(y,0)\right\rangle \right]\right\} \end{eqnarray*}
and a similar relation for the left fermion. It then suffices to insert
the expressions derived in the previous section.

For instance at zero temperature:\begin{eqnarray*}
 & \left\langle R(x,t)R^{+}(y,0)\right\rangle \\
= & \frac{1}{2\pi\delta}\,\prod_{n=-\infty}^{+\infty}\left[\frac{\left(\delta+i2nL\right)^{2}}{\left[\delta+i(ut+2nL)\right]^{2}+(x-y)^{2}}\right]^{\frac{K+K^{-1}}{4}(r_{S}r_{D})^{\left|n\right|}}\\
\times & \prod_{n=-\infty}^{+\infty}\left[\frac{\delta+i(ut+2nL+x-y)}{\delta+i(ut+2nL+y-x)}\right]^{\frac{(r_{S}r_{D})^{\left|n\right|}}{2}}\\
\times & \prod_{\varepsilon=\pm1,\, n=0}^{+\infty}\left[\frac{\left[\delta+i(ut-x-y+\varepsilon(2n+1)L)\right]^{2}}{\left[\delta+i(ut-2x+\varepsilon(2n+1)L)\right]\left[\delta+i(ut-2y+\varepsilon(2n+1)L)\right]}\right]^{\frac{K-K^{-1}}{8}(r_{S}r_{D})^{\left|n\right|}r_{S}}\\
\times & \prod_{\varepsilon=\pm1,\, n=0}^{+\infty}\left[\frac{\left[\delta+i(ut+x+y+\varepsilon(2n+1)L)\right]^{2}}{\left[\delta+i(ut+2x+\varepsilon(2n+1)L)\right]\left[\delta+i(ut+2y+\varepsilon(2n+1)L)\right]}\right]^{\frac{K-K^{-1}}{8}(r_{S}r_{D})^{\left|n\right|}r_{D}}\end{eqnarray*}
and appropriate expressions at finite-temperature. The tunneling DOS
is the Fourier transform of this correlator.

\section{Conclusion.}

We have extended the bosonization technique to a LL connected to resistances
computing correlators of the boson fields in so doing. The latter
are the building blocks allowing calculation of the fermion correlators.
As side results we derived also the finite-frequency conductivity
and found that the finite-size LL with IBC is equivalent to an infinite
LL by virtue of identity of Green's functions whenever impedance matching
is realized. 

We also recovered explicitly the properties of an open LL: it corresponds
to IBC with infinite resistances. 

But in general the LL with IBC has distinctly different properties
(it is a dissipative system) and forms a universality class in its
own right much as the LL with OBC which exhibits critical exponents
different from those of the infinite LL. Further interesting developments
using the results in this paper would be a study of the single particle
spectral density which is the object of interest in tunneling experiments.
A study of the shot noise using the Keldysh technique would also be
straightforward given the knowledge of the Green's functions.

\section*{Appendix A: Conductivity.\label{sec:Appendix-A}}

We use linear response theory: let us consider the perturbation $\hat{V}=-\int dx\rho\: U$
where $U$ is a voltage. Integrating by parts one gets: $\hat{V}=-\int dx\, E\:\frac{\phi}{\sqrt{\pi}}$.
By linear response:\[
\delta\left\langle \phi(x,t)\right\rangle =\frac{i}{\sqrt{\pi}}\int_{0}^{L}dy\int_{-\infty}^{+\infty}\, dt'\, G_{R}(x,t;y,t')\, E(y,t').\]

But $i_{e}(t)=-\frac{e}{\sqrt{\pi}}\:\frac{\partial\phi}{\partial t}$
therefore:\[
i_{e}(x,\omega)=\frac{i\, e\,\omega}{\sqrt{\pi}}\,\phi(x,\omega)=-\frac{e}{\pi}\,\omega\,\int dy\, G_{R}(x,y,\omega)\, E(y,\omega)\]
 so that the non-local conductivity is: \[
\sigma(x,y,\omega)=\frac{e}{\pi}\,\omega\, G_{R}(x,y,\omega).\]

The expression of $G_{R}(x,y,\omega)$ was computed in section \ref{sec:Green's-function-of}.
This expression allows measurement of the complex boundary impedances.

\section*{Appendix B: Recovering the Luttinger Liquid with open boundary conditions.\label{sec:Appendix-B}}

In this Appendix we will show that the LL with OBC is a special limit
of the LL with IBC when the reflection parameters at the boundaries
are set to unity. Physically this comes about because perfect reflection
can be equated with having a hard-wall.

The OBC for the fermion operator:\[
\psi(x)=0\;\;(x=0\; or\; L)\]
i.e. \begin{eqnarray*}
\psi_{R}(0) & = & -\psi_{L}(0)\\
\psi_{R}(L) & = & -\psi_{L}(L)\end{eqnarray*}
will be derived explicitly from the IBC (with $r_{S}=1=r_{D}$) (we
have dropped the phase $e^{i2k_{F}L}$ since $k_{F}=N\pi/L$). This
will show the perfect equivalence between the LL with OBC and the
LL with IBC when $r_{S}=1=r_{D}$. But already we observe that:

\[
OBC\;\Longrightarrow IBC\;(r_{S}=1=r_{D})\]
since cancellation of the fermion operator implies that its current
is also zero (and particularly its $k\ll k_{F}$ harmonics: see for
instance the relation $\rho_{+}(x)=\rho_{-}(-x)$ found for the OBC,
third equation below eq.(10) of Ref. 4) ).

But for reflection coefficients $r_{S}=1=r_{D}$ the IBC is equivalent
to stating that $\rho_{+}=\rho_{-}$ at both boundaries (therefore
the current $I\propto\rho_{+}-\rho_{-}$vanishes). The OBC therefore
does imply the IBC. 

We now prove the converse and establish:\[
IBC\;(r_{S}=1=r_{D})\;\Longrightarrow OBC.\]

The proof actually exists already at $90\%$ in Ref. 5 which in order
to quantize the LL with OBC actually started from the zero current
condition on the boson field: in that work the fermionic boundary
conditions are used to derive the quantization rules on charges. But
one can take a reverse standpoint: this allows to derive the fermionic
boundary conditions by demanding that charge be conserved. (We will
also in the course of the proof reconcile Ref. 4 and Ref. 5 which
find a slightly different Green's function {[}see eq.(29) of Ref.
5 and the discussion after it about the difference with eq.(31) of
Ref. 4] .)

\subsection{Derivation of the OBC.}

\subsubsection{Mode development of the fields.}

We depart in this Appendix from the definition of the boundaries at
$x=\pm L/2$ to set them at $x=0,\: L$. For reflection coefficients
$r_{S}=1=r_{D}$ the IBC imply:\begin{eqnarray*}
\partial_{x}\phi_{+}(0,t) & = & \partial_{x}\phi_{-}(0,t),\\
\partial_{x}\phi_{-}(L,t) & = & \partial_{x}\phi_{+}(L,t).\end{eqnarray*}

Therefore:\begin{eqnarray*}
\partial_{t}\phi_{+}(0,t) & = & -\partial_{t}\phi_{-}(0,t),\\
\partial_{t}\phi_{-}(L,t) & = & -\partial_{t}\phi_{+}(L,t).\end{eqnarray*}
This implies:\begin{equation}
\phi(0,t)=C_{0},\;\;\phi(L,t)=C_{L}\label{eq:bcphi}\end{equation}
 where the operators $C_{0}$ and $C_{L}$ do not depend on time.

For these values of $r_{S}$ and $r_{D}$ the boson theory is dissipationless
and will be described by a $c=1$ conformal field theory. It will
be necessary to make an eigenmode development of the fields. To do
it we follow closely Ref. 5 who treated the quantization of the OBC
by using as a starting point eq. (\ref{eq:bcphi}) (i.e. by using
the IBC!). Since the field $\phi$ obeys the standard wave equation
and given the boundary conditions ( eq.(\ref{eq:bcphi}) ) the mode
expansion for $\phi$ and $\Theta$ must have the form:\begin{eqnarray}
\phi(x,t) & = & \phi_{0}+\frac{\sqrt{\pi}}{L}\, Q\: x+\sqrt{K}\sum_{n\geq1}\frac{\sin(q_{n}x)}{\sqrt{\pi n}}\left(-ia_{n}e^{-iq_{n}ut}+ia_{n}^{+}e^{iq_{n}ut}\right)\label{eq:modexp}\\
\Theta(x,t) & = & \Theta_{0}-\frac{\sqrt{\pi}}{L}\,\frac{Q}{K}\: ut-\frac{1}{\sqrt{K}}\sum_{n\geq1}\frac{\cos(q_{n}x)}{\sqrt{\pi n}}\left(a_{n}e^{-iq_{n}ut}+a_{n}^{+}e^{iq_{n}ut}\right)\nonumber \end{eqnarray}
 where $q_{n}=n\pi/L$.

We impose the standard equal time commutation relations for the fields:\[
\left[\phi(x);\:\phi(y)\right]=0,\;\left[\Theta(x);\:\Theta(y)\right]=0,\;\left[\phi(x);\:\Theta(y)\right]=i\theta(x-y)\]
where $\theta(x)$ is the Heaviside step function. The first two commutators
$\left[\phi(x);\:\phi(y)\right]=0,\;\left[\Theta(x);\:\Theta(y)\right]=0$
imply that all the commutators vanish except $\left[a_{n};\, a_{n}^{+}\right]$
and $\left[Q;\:\Theta_{0}\right]$. Expanding the third commutator
one gets:\[
i\theta(x-y)=\frac{\sqrt{\pi}}{L}\,\left[Q;\:\Theta_{0}\right]x+2i\sum_{n\geq1}\frac{\sin(q_{n}x)\cos(q_{n}y)}{\pi n}\left[a_{n};\: a_{n}^{+}\right]\]
 and by making use of the expansion:\[
i\theta(x-y)=i\frac{x}{L}+2i\sum_{n\geq1}\frac{\sin(q_{n}x)\cos(q_{n}y)}{\pi n}\]
which can be proved by using eq. (A-1) of Ref. 4 one finds that the
only non-zero commutators are:\[
\left[a_{n};\: a_{n}^{+}\right]=1,\;\left[Q;\:\Theta_{0}\right]=i/\sqrt{\pi}.\]
We note that $\phi_{0}$ is a c-number; this is normal since the operator
$J$ which is usually its conjugate momentum in the periodic LL does
not appear in the theory. We can therefore remove it altogether since
the LL Hamiltonian is invariant under constant shifts of the phase
field: $\phi\longrightarrow\phi-\phi_{0}$.

The zero mode $Q$ is as usual the charge added to the system since
the number density is $\rho-\rho_{0}=\frac{1}{\sqrt{\pi}}\partial_{x}\phi$:\begin{eqnarray*}
N-N_{0} & = & \frac{1}{\sqrt{\pi}}\left(\phi(L)-\phi(0)\right),\\
 & = & Q.\end{eqnarray*}

\subsubsection{Picking the right bosonization formula.}

The usual bosonization formula for the fermion operator used in infinite
systems \[
\psi_{R/L}=\frac{1}{\sqrt{2\pi\delta}}\:\exp i\sqrt{\pi}\left(\pm\phi+\Theta\right)\]
 is not the only one possible; other valid vertex operators for a
fermion operator are for instance:

\begin{eqnarray*}
\widetilde{\psi}_{R} & = & \frac{e^{i\alpha x+\beta}}{\sqrt{2\pi\delta}}\:\exp i\sqrt{\pi}\left(\phi+\Theta\right),\\
\widetilde{\psi}_{L} & = & \frac{e^{-i\alpha x+\gamma}}{\sqrt{2\pi\delta}}\:\exp i\sqrt{\pi}\left(\phi+\Theta\right)\end{eqnarray*}
where $\alpha,\beta$ and $\gamma$ are real constants. It is easy
to check that these constants do not affect the bosonization proof
given in Section 5: $\beta$ and $\gamma$ correspond to a $U(1)_{L}\times U(1)_{R}$
invariance of the free Dirac lagrangian while $\alpha$ only shifts
the chemical potential (substitution in the Dirac Hamiltonian leads
simply to the additional term $\alpha(N_{R}+N_{L})$). The effect
of these constants on the physics is slight: the $\alpha$ term leads
to additional oscillations in the Green's function, which are in a
sense trivial because they only correspond to a shift in chemical
potential; the other constants have no effect. However when one wishes
the operators to obey specific boundary conditions they will be necessary.

Indeed this liberty in the choice of the fermion operator is useful:
remember that OBC can be obtained from a variety of conditions, for
instance Dirichlet BC (vanishing of the fermionic wavefunction) or
Neumann BC (vanishing of its derivative). Dirichlet BC $\psi=0$ at
a boundary at $x=0$ or $x=L$ reads:\begin{eqnarray*}
\psi_{R}(0) & = & -\psi_{L}(0),\\
\psi_{R}(L) & = & -e^{-i2k_{F}L}\psi_{L}(L).\end{eqnarray*}
 A Neumann BC reads $\partial_{x}\psi=0$. In the low-energy limit
$k\ll k_{F}$: \[
\partial_{x}\psi=0\;\Longrightarrow\; k_{F}\left(e^{ik_{F}x}\psi_{R}-e^{-ik_{F}x}\psi_{L}\right)=0\]
and therefore at $x=0,\; L$: \begin{eqnarray*}
\psi_{R}(0) & = & \psi_{L}(0),\\
\psi_{R}(L) & = & e^{-i2k_{F}L}\psi_{L}(L).\end{eqnarray*}

Therefore the same vertex operator can not obey both Dirichlet and
Neumann BC at the same time (at the same location). We now give the
correct prescriptions for both situations and also mixed ones (Dirichlet
at one boundary and Neumann at the other). 

It will be convenient to define in the following discussion the 'primed'
operators which differ from the usual ones by a shift of the Fermi
vector $\frac{\pi}{2L}$:

\begin{eqnarray*}
\psi'_{R}(x,t) & = & \frac{1}{\sqrt{2\pi\delta}}\:\exp i\sqrt{\pi}\left(\phi+\Theta\right)\:\exp i\frac{\pi}{2L}(x-\frac{u}{K}t);\\
\psi'_{L}(x,t) & = & \frac{1}{\sqrt{2\pi\delta}}\:\exp i\sqrt{\pi}\left(-\phi+\Theta\right)\:\exp-i\frac{\pi}{2L}(x+\frac{u}{K}t).\end{eqnarray*}
These operators correspond actually to factoring out the zero modes
from the phase fields: 

\begin{eqnarray*}
\psi'_{R}(x,t) & = & \frac{1}{\sqrt{2\pi\delta}}\:\exp i\sqrt{\pi}\Theta_{0}\:\exp i\pi\frac{Q}{L}(x-\frac{u}{K}t)\:\exp i\sqrt{\pi}\left(\phi'+\Theta'\right)\\
\psi'_{L}(x,t) & = & \frac{1}{\sqrt{2\pi\delta}}\:\exp i\sqrt{\pi}\Theta_{0}\:\exp-i\pi\frac{Q}{L}(x+\frac{u}{K}t)\:\exp i\sqrt{\pi}\left(-\phi'+\Theta'\right)\end{eqnarray*}
(where the prime ' means fields from which the zero modes have been
subtracted). These relations follow immediately from Campbell-Haussdorf
formula and the commutator $\left[Q;\:\Theta_{0}\right]=i/\sqrt{\pi}$
($\phi'$ and $\Theta'$ commute with the zero modes).

\textbf{Dirichlet boundary conditions.}

In the free-fermion limit the zeros at the two boundaries imply that
the Fermi vector has the form: $k_{F}L=\pi N$ so we can remove the
phase $e^{-i2k_{F}L}$ from the BC.

Let us consider the operators:

\begin{eqnarray*}
\psi_{R}^{D}(x,t) & = & \psi'_{R}(x,t);\\
\psi_{L}^{D}(x,t) & =- & \psi'_{L}(x,t).\end{eqnarray*}
(the upper index $D$ stands for Dirichlet).

We now use the mode expansion on the fields (eq.(\ref{eq:modexp})
) which implies: \[
\phi'(0)=0=\phi'(L).\]

This in turn implies: \begin{eqnarray*}
\psi_{R}^{D}(0,t) & = & \frac{1}{\sqrt{2\pi\delta}}\:\exp i\sqrt{\pi}\Theta_{0}\:\exp-i\pi\frac{Q}{L}\frac{u}{K}t\:\exp i\sqrt{\pi}\Theta'(0,t)\\
\psi_{L}^{D}(0,t) & = & \frac{-1}{\sqrt{2\pi\delta}}\:\exp i\sqrt{\pi}\Theta_{0}\:\exp-i\pi\frac{Q}{L}\frac{u}{K}t\:\exp i\sqrt{\pi}\Theta'(0,t)\end{eqnarray*}
 Therefore at $x=0$ we have trivially a Dirichlet BC:\[
\psi_{R}^{D}(0,t)=-\psi_{L}^{D}(0,t).\]

At $x=L$: \begin{eqnarray*}
\psi_{R}(L,t) & =\frac{1}{\sqrt{2\pi\delta}}\: & \exp i\sqrt{\pi}\Theta_{0}\:\exp i\pi Q\:\exp i\pi\frac{Q}{KL}(-ut)\:\exp i\sqrt{\pi}\Theta',\\
\psi_{L}^{D}(L,t) & =\frac{-1}{\sqrt{2\pi\delta}}\: & \exp i\sqrt{\pi}\Theta_{0}\:\exp-i\pi Q\:\exp i\pi\frac{Q}{KL}(-ut)\:\exp i\sqrt{\pi}\Theta'.\end{eqnarray*}
The two expressions are almost identical except for the second term
at the right of the sign equal. We now enforce charge quantization:
the operator $Q$ must have only integral eigenvalues. Therefore we
have the \emph{operator} equality:\[
\exp i\pi Q=\exp-i\pi Q.\]
 This then implies Dirichlet BC at $x=L$ (with $k_{F}L=\pi N$ ):\[
\psi_{R}^{D}(L,t)=-\psi_{L}^{D}(L,t).\]

Furthermore using the mode expansion one finds that: \[
\phi'(x)+\Theta'(x)=-\phi'(-x)+\Theta'(-x).\]
Since: \begin{eqnarray*}
\psi_{R}^{D} & = & \frac{1}{\sqrt{2\pi\delta}}\:\exp i\sqrt{\pi}\Theta_{0}\:\exp i\pi\frac{Q}{L}(x-\frac{ut}{K})\:\exp i\sqrt{\pi}\left(\phi'+\Theta'\right)\\
\psi_{L}^{D} & = & -\frac{1}{\sqrt{2\pi\delta}}\:\exp i\sqrt{\pi}\Theta_{0}\:\exp i\pi\frac{Q}{L}(-x-\frac{ut}{K})\:\exp i\sqrt{\pi}\left(-\phi'+\Theta'\right)\end{eqnarray*}
it follows that:\[
\psi_{R}^{D}(x)=-\psi_{L}^{D}(-x).\]

One can check that the prescription is exactly that of F-G in Ref.
4 (see eq. (9) and eq. (7)).

We have thus recovered Dirichlet boundary conditions starting from
the IBC at $r_{S}=1=r_{D}$.

\textbf{Neumann boundary conditions. }

Again in the free-fermion limit the zeros at the two boundaries imply
that the Fermi vector has the form: $k_{F}L=\pi N$ so we can remove
the phase $e^{-i2k_{F}L}$ from the BC.

Since we are at liberty to add constant phases and still get vertex
operators for fermions, the previous discussion suggests that suitable
expressions are:\begin{eqnarray*}
\psi_{R}^{N} & = & \psi'_{R},\\
\psi_{L}^{N} & = & \psi'_{L}.\end{eqnarray*}

This leads immediately to Neumann BC at $x=0$ and $x=L$:

\begin{eqnarray*}
\psi_{R}^{N}(0) & = & \psi_{L}^{N}(0),\\
\psi_{R}^{N}(L) & = & \psi_{L}^{N}(L).\end{eqnarray*}

\textbf{Mixed boundary conditions.}

For mixed boundary conditions $k_{F}$ is quantized as $k_{F}L=\pi(N+\frac{1}{2})$
(which is the correct quantization for free fermions in a box with
mixed conditions $\psi=0$ at one end and $\partial_{x}\psi=0$ at
the other one).

- for Dirichlet BC at $x=0$ and Neumann BC at $x=L$: \begin{eqnarray*}
\psi_{R}(0) & = & -\psi_{L}(0)\\
\psi_{R}(L) & = & e^{-i2k_{F}L}\psi_{L}(L)=-\psi_{L}(L).\end{eqnarray*}
therefore the expressions of the fermion fields used for Dirichlet
BC still work: \begin{eqnarray*}
\psi_{R}^{D}(x,t) & = & \psi'_{R}\\
\psi_{L}^{D}(x,t) & = & -\psi'_{L}\end{eqnarray*}

- for Neumann BC at $x=0$ and Dirichlet BC at $x=L$, the Neumann
prescription above is the correct one.

\textbf{A 'Twisted' boundary condition.}

The interested reader may inquire what boundary conditions the standard
operators $\psi_{R/L}$ can describe; it can be checked that:\[
\psi_{R/L}=\frac{1}{\sqrt{2\pi\delta}}\:\exp i\sqrt{\pi}\left(\pm\phi+\Theta\right)\]
implies:\begin{eqnarray*}
\psi_{R}(0) & = & \psi_{L}(0),\\
\psi_{R}(L) & =- & \psi_{L}(L).\end{eqnarray*}
This can describe a system with any combination of Dirichlet or Neumann
BC provided we add the condition that one boundary adds a $\pi$ phase;
a way to do that with free fermions is to add a boundary interaction
with reflection coefficient $r=-1$ such as $\psi_{R}^{+}\psi_{R}+\psi_{L}^{+}\psi_{L}-\psi_{R}^{+}\psi_{L}-\psi_{L}^{+}\psi_{R}$
. If we unfold the non-chiral system of length $L$ into a chiral
system of length $2L$ this corresponds to the theory of a single
chiral fermion on a circle threaded by a flux $\pi$.

\subsubsection{Fabrizio-Gogolin Bosonization of OBC versus Mattsson et coll. Bosonization.}

The earliest theory on the LL with OBC is due to Fabrizio and Gogolin
(F-G)\cite{key-4}. A little bit later the problem was also treated
by Mattsson and collaborators (M-E-J)\cite{key-5} with a \emph{different
bosonization scheme} namely the standard prescription used for the
infinite system: \[
\psi_{R/L}=\frac{1}{\sqrt{2\pi\delta}}\:\exp i\sqrt{\pi}\left(\pm\phi+\Theta\right).\]
The treatments lead mostly to the same results although there are
in details some minor differences. For instance, M-E-J find that the
Green's function has an additional modulation coming from zero modes
not present in F-G's results (see eq. (29) of Ref. 5 and the discussion
which follows and compare to eq. (31) of Ref. 4):\[
G_{M-E-J}(x,t;y,0)=e^{-i\frac{\pi}{2L}(2n_{0}-1)(x-y+u_{c}K_{c}^{-1}t)}G_{F-G}(x,t;y,0)\]
(the Green's function written is that of the left fermion $\psi_{L}$
with $n_{0}$ defined by M-E-J as $n_{0}=k_{F}L/\pi\: mod\,1$). M-E-J
comment that this difference with F-G will have implications for time
correlations but do not explain the origin of the discrepancy.

The previous discussion should hint at the explanation: the additional
phase as compared with F-G comes from the fact that F-G use the primed
operators $\psi'_{R/L}$. 

This leads naturally to the question: which is the correct prescription
since M-E-J also use Dirichlet BC? We now show that M-E-J choice leads
to inconsistencies with regard to conserved charges and that the correct
prescription for Dirichlet BC is indeed that of F-G.

M-E-J find (eq.(22a-b) of Ref. 5 and eq.(5) and (18) for definitions)
for the total charge $Q_{c}=Q_{\uparrow}+Q_{\downarrow}$and $Q_{s}=Q_{\uparrow}-Q_{\downarrow}$(after
proper rescaling to extract the physical charges):\begin{eqnarray*}
Q_{c} & = & n+1+\frac{2k_{F}L}{\pi}\\
Q_{s} & = & m\end{eqnarray*}
 where $n$ and $m$ are integers having the same parity. M-E-J derived
these constraints by imposing the Dirichlet BC. 

But for free fermions with Dirichlet BC at both boundaries: $k_{F}L=\pi N$
(which implies that $\frac{2k_{F}L}{\pi}$ is an even number!). M-E-J
equations therefore imply that $Q_{c}$ and $Q_{s}$ have opposite
parity.

However since $Q_{c}=Q_{s}+2Q_{\downarrow}$, $Q_{c}$ and $Q_{s}$
must have the same parity: we have therefore a contradiction. Having
the correct quantization conditions plays an important role for the
partition function and for the finite temperature Green's function
for the zero modes part so the issue is not innocuous.

An obvious way to cure the problem would be to artificially prescribe
that the Fermi wavevector is $k'_{F}L=\pi(N+\frac{1}{2})$ since this
avoids the problem with charges: it also removes the additional phase
M-E-J find in the Green's function and yields Dirichlet BC but of
course the value of $k'_{F}$ is incorrect in the limit of free fermions.
A possible interpretation of the shift might be that it proceeds from
a change in Maslov-Morse index in the trajectory.

\subsection{Recovering the correlators of OBC.}

We will check here directly using the expressions computed with IBC
that we recover the correlators of OBC. (NB: In order to compare our
results with those of the litterature which work with $x\in\left[0,L\right]$,
since we worked with $x\in\left[-L/2,L/2\right]$ we shift all the
space arguments by $L/2$ when using the expressions derived in the
bulk of this paper.)

\subsubsection{Boson correlator.}

\textbf{\textit{\emph{Chiral boson correlator:}}}

For $r_{S}=1=r_{D}$ the chiral boson fields correlators read:

\[
F_{++}(x,t;\: y,0)=\left\langle \phi_{+}(x,t)\phi_{+}(y,0)\right\rangle =-\frac{K}{4\pi}\sum_{n=-\infty}^{+\infty}\left(r_{S}r_{D}\right)^{\left|n\right|}\ln\left[\delta+i(ut-2nL-x+y)\right].\]
One can also check directly that:\begin{eqnarray*}
\left\langle \phi_{+}(x,t)\phi_{-}(y,0)\right\rangle  & =- & F_{++}(x,t;\:-y,0),\\
\left\langle \phi_{-}(x,t)\phi_{+}(y,0)\right\rangle  & =- & F_{++}(-x,t;\: y,0),\\
\left\langle \phi_{-}(x,t)\phi_{-}(y,0)\right\rangle  & = & F_{++}(-x,t;\:-y,0).\end{eqnarray*}

Using the product expansion of the sine function: \[
\sin(\pi z)=\pi z\prod_{n\geq1}(1-\frac{z^{2}}{n^{2}})\]
one gets:\[
\Delta F_{++}=F_{++}(x,t;\: y,0)-F_{++}(0,0;\:0,0)=-\frac{K}{4\pi}\ln\left[\frac{2L}{\pi\delta}\frac{\delta+i(ut-x+y)}{ut-x+y}\sin\frac{\pi(ut-x+y)}{2L}\right].\]
This yields:\begin{eqnarray*}
{\displaystyle \Delta F_{++}} & = & \left\{ \begin{array}{cc}
-\frac{K}{4\pi}\ln\left[\frac{i2L}{\pi\delta}\sin\frac{\pi(ut-x+y)}{2L}\right] & \: for\: ut-x+y\neq0;\\
0 & \: for\: ut-x+y=0.\end{array}\right.\end{eqnarray*}
This agrees with the expressions found in the litterature (e.g. eq.
(28) in Ref. 5) in the limit of zero temperature for finite-length.

One checks easily that \[
{\displaystyle \Delta F_{++}}=\frac{K}{4\pi}\left\{ i\pi\frac{x-y-ut}{2L}-S(x-y-ut)+S(0)\right\} \]
where $S(z)$ introduced by F-G is (see eq.(A-1) of Ref. 4):\begin{eqnarray*}
S(z) & = & \ln\frac{L}{\pi\delta}P(z)+if(z)\\
P(z) & = & \frac{\pi\delta}{2L\sqrt{\sinh^{2}\frac{\pi\delta}{2L}+\sin^{2}\frac{\pi z}{2L}}}\\
f(z) & = & \arctan\frac{\sin\frac{\pi z}{2L}}{\exp\frac{\pi\delta}{L}-\cos\frac{\pi z}{2L}}.\end{eqnarray*}
This cumbersome expression will be useful to neatly separate the phase
in the fermion Green's function.

\textbf{Non-chiral correlator:}

Using our previous results (eq. (\ref{eq:phicorrel}) ) one has (we
keep the calculations at zero temperature for simplicity):

\begin{eqnarray*}
\left\langle \phi(x,t)\phi(y,0)\right\rangle  & = & -\frac{K}{4\pi}\;\sum_{n=-\infty}^{+\infty}\ln\left\{ \left[\delta+i(ut+2nL)\right]^{2}+(x-y)^{2}\right\} \\
 &  & +\frac{K}{4\pi}\;\sum_{n=-\infty}^{+\infty}\ln\left[\delta+i(ut-x-y+(2n+1)L)\right]\\
 &  & +\frac{K}{4\pi}\;\sum_{n=-\infty}^{+\infty}\ln\left[\delta+i(ut+x+y+(2n+1)L)\right].\end{eqnarray*}

We regularize by adding $-\frac{1}{2}\left\langle \phi(x,t)^{2}\right\rangle -\frac{1}{2}\left\langle \phi(y,0)^{2}\right\rangle $
and by using the infinite product expressions for $\cos$ and $\sin$
($\cos(\frac{\pi}{2}z)=\prod_{n\geq1}(1-\frac{z^{2}}{(2n+1)^{2}})$
and $\sin(\pi z)=\pi z\prod_{n\geq1}(1-\frac{z^{2}}{n{}^{2}})$ )
one gets:\begin{eqnarray*}
 & \left\langle \phi(x,t)\phi(y,0)\right\rangle -\frac{1}{2}\left\langle \phi(x,t)^{2}\right\rangle -\frac{1}{2}\left\langle \phi(y,0)^{2}\right\rangle \\
=-\frac{K}{4\pi} & \ln\left\{ \left(\frac{2L}{\pi u\delta}\right)^{2}\sin\left(\frac{\pi}{2L}(ut-x+y)\right)\sin\left(\frac{\pi}{2L}(ut+x-y)\right)\right\} \\
+\frac{K}{4\pi} & \ln\left\{ \left(\frac{\cos\left(\frac{\pi}{2L}(ut+x+y)\right)\cos\left(\frac{\pi}{2L}(ut-x-y)\right)}{\cos\left(\frac{\pi}{L}x\right)\cos\left(\frac{\pi}{L}y\right)}\right)\right\} .\end{eqnarray*}
Ref. \cite{key-5} does not give explicitly the full correlator for
the field $\phi$ and only the following chiral correlator for the
left field $\phi_{L}$ is derived for $K=1$:\begin{eqnarray*}
F_{--}^{M-E-J} & = & \left\langle \phi_{L}(x,t)\phi_{L}(y,0)\right\rangle -\frac{1}{2}\left\langle \phi_{L}(x,t)^{2}\right\rangle -\frac{1}{2}\left\langle \phi_{L}(y,0)^{2}\right\rangle \\
 & = & -\frac{K}{4\pi}\;\ln\left\{ \left(\frac{2L}{\pi u\delta}\right)\sin\left(\frac{\pi}{2L}(ut+x-y)\right)\right\} \end{eqnarray*}
For the sake of comparison with our expression let us rebuild the
full correlator using the previous equation. Indeed taking into account
the OBC leads to: \begin{eqnarray*}
F_{++}(x,t;y,0) & = & F_{--}(-x,t;-y,0)\\
F_{+-}(x,t;y,0) & =- & F_{--}(-x,t;y,0)\\
F_{-+}(x,t;y,0) & =- & F_{--}(x,t;-y,0)\end{eqnarray*}
which yields finally (after rescaling the fields to make the LL parameter$K$
appear): \begin{eqnarray*}
F^{M-E-J}= & \left\langle \phi(x,t)\phi(y,0)\right\rangle -\frac{1}{2}\left\langle \phi(x,t)^{2}\right\rangle -\frac{1}{2}\left\langle \phi(y,0)^{2}\right\rangle \\
=-\frac{K}{4\pi} & \ln\left\{ \left(\frac{2L}{\pi u\delta}\right)^{2}\sin\left(\frac{\pi}{2L}(ut-x+y)\right)\sin\left(\frac{\pi}{2L}(ut+x-y)\right)\right\} \\
+\frac{K}{4\pi} & \ln\left\{ \left(\frac{\sin\left(\frac{\pi}{2L}(ut+x+y)\right)\sin\left(\frac{\pi}{2L}(ut-x-y)\right)}{\sin\left(\frac{\pi}{L}x\right)\sin\left(\frac{\pi}{L}y\right)}\right)\right\} .\end{eqnarray*}
This is identical with our result if one takes care to shift the origin
to the left boundary $x\rightarrow x+L/2$ (since Ref. \cite{key-5}
uses the left boundary as origin). 

Our results are therefore in perfect agreement with the calculations
of Ref. 4 and Ref. 5.

\subsubsection{Fermion correlator.}

After recovering the boson correlator we turn to the fermion operator
for a spinful LL (with customary definitions):

\[
\psi_{R\uparrow}=\frac{1}{\sqrt{2\pi\delta}}\:\exp i\sqrt{\pi}\left(\phi_{\uparrow}+\Theta_{\uparrow}\right)\;\exp i\frac{\pi}{2L}(x-\sum_{\nu=c,s}\frac{u_{\nu}}{2K_{\nu}}t)\]
 where the phase comes from the fact that the operators obeying the
Dirichlet boundary conditions have the zero modes extracted from the
exponential as explained above.

Defining the correlator as:\[
G_{R\uparrow}=-i\left\langle \psi_{R\uparrow}(x,t)\psi_{R\uparrow}^{+}(y,0)\right\rangle \]
and inserting the expressions of the (charge and spin) chiral fields
one finds:\[
G_{R\uparrow}=-\frac{i}{2\pi\delta}\exp i\frac{\pi}{2L}(x-y-\sum_{\nu=c,s}\frac{u_{\nu}}{2K_{\nu}}t)\:\exp\frac{\pi}{2}\sum_{\nu=c,s}C_{\nu}(x,t;\: y)\]
where:\begin{eqnarray*}
C_{\nu}(x,t;\: y) & = & (1+K_{\nu}^{-1})^{2}\Delta F_{++}^{\nu}(x,t;\: y)+(1-K_{\nu}^{-1})^{2}\Delta F_{++}^{\nu}(-x,t;\:-y)\\
- & (1-K_{\nu}^{-2}) & \left\{ F_{++}^{\nu}(-x,t;\: y)+F_{++}^{\nu}(x,t;\:-y)\right.\\
 & -\frac{1}{2} & \left.\left[F_{++}^{\nu}(x,0;\, x)+F_{++}^{\nu}(-x,0;\,-x)+F_{++}^{\nu}(y,0;\, y)+F_{++}^{\nu}(-y,0;\,-y)\right]\right\} \end{eqnarray*}

Substitution of the chiral boson correlator yields as a function of
$P(z)$ and $f(z)$:

\begin{eqnarray*}
G_{R\uparrow}=\frac{-i}{2\pi\delta}\left(\frac{L}{\pi\delta}\right)^{\frac{1}{4}\sum_{\nu}K_{\nu}+K_{\nu}^{-1}}\prod_{\nu=c,s}P(x-y-u_{\nu}t)^{c_{\nu}^{2}/2}P(x-y+u_{\nu}t)^{s_{\nu}^{2}/2}\\
\times\left[\frac{P(x+y-u_{\nu}t)P(x+y+u_{\nu}t)}{P(2x)P(2y)}\right]^{-s_{\nu}c_{\nu}/2}\exp i\Phi(x,y,t)\end{eqnarray*}
where if we define $K_{\nu}=\exp2\chi_{\nu}$:\[
c_{v}=\cosh\chi_{\nu},\; s_{\nu}=\sinh\chi_{\nu}\]
and where the phase factor $\Phi(x,y,t)$ (it is not the boson field
$\phi$!) is given by:

\begin{eqnarray*}
\Phi(x,y,t) & = & \frac{\pi}{L}(x-y)-\frac{\pi}{2L}\sum_{\nu}\frac{u_{\nu}}{K_{\nu}}t+\frac{1}{2}\sum_{\nu}\left[c_{\nu}^{2}f(x-y-u_{\nu}t)\right.\\
 & - & \left.s_{\nu}^{2}f(x-y+u_{\nu}t)+s_{\nu}c_{\nu}\left(f(x+y+u_{\nu}t)-f(x+y-u_{\nu}t)\right)\right].\end{eqnarray*}

Our expression coincides with that found by F-G (eq. (31) in Ref.
4) apart from two things: (i) an unessential constant $\left(\frac{L}{\pi\delta}\right)^{\frac{1}{4}\sum_{\nu}K_{\nu}+K_{\nu}^{-1}}$which
comes from a difference of normalization of the Fermion operator (eq.
(9) of Ref. 4 has a prefactor $\propto1/\sqrt{2L}$ while we have
$\propto1/\sqrt{2\pi\delta}$); (ii) the phase factors $\Phi(x,y,t)$
are identical except for the term $\frac{\pi}{\mathbf{2}L}\sum_{\nu}\frac{u_{\nu}}{K_{\nu}}t$
(F-G have $\frac{\pi}{\mathbf{4}L}\sum_{\nu}\frac{u_{\nu}}{K_{\nu}}t$,
{[}factor 2 instead of 4] ). 

One can check however by taking the non-interacting limit that the
term $\frac{\pi}{2L}\sum_{\nu}\frac{u_{\nu}}{K_{\nu}}t$ is the correct
one: in the non-interacting case the phase must be a chiral function
of $x-y-v_{F}t$ since we compute a right fermion correlator; substituting
$K_{\nu}=1$, $s_{\nu}=0$ and $c_{\nu}=1$ , $u_{\nu}=v_{F}$ in
our expression then yields $\Phi(x,y,t)=\frac{\pi}{L}(x-y-v_{F}t)+f(x-y-v_{F}t)$
which has the right dependence (which F-G can not have with the $\frac{\pi}{4L}\sum_{\nu}\frac{u_{\nu}}{K_{\nu}}t$
term).

In conclusion we have recovered using IBC the fermion correlator for
OBC (correcting in passing a misprint in Ref. 4).

\end{document}